\documentclass[final,3p,authoryear,12pt]{elsarticle}

\usepackage{pslatex}

\usepackage{amssymb,graphicx,epsfig}

\journal{New Astronomy Reviews}

\begin{document}

\begin{frontmatter}



\title{Circumstellar Debris and Pollution at White Dwarf Stars\footnote{Dedicated to the memory of Prof.\ M.\ Jura (1947--2016).}}


\author{J. Farihi}

\address{Department of Physics \& Astronomy, University College London\\j.farihi@ucl.ac.uk}

\begin{abstract}

Circumstellar disks of planetary debris are now known or suspected to closely orbit hundreds of white dwarf stars.  To date, both data and 
theory support disks that are entirely contained within the preceding giant stellar radii, and hence must have been produced during the white 
dwarf phase.  This picture is strengthened by the signature of material falling onto the pristine stellar surfaces; disks are always detected 
together with atmospheric heavy elements.  The physical link between this debris and the white dwarf host abundances enables unique insight 
into the bulk chemistry of extrasolar planetary systems via their remnants.  This review summarizes the body of evidence supporting dynamically 
active planetary systems at a large fraction of all white dwarfs, the remnants of first generation, main-sequence planetary systems, and hence 
provide insight into initial conditions as well as long-term dynamics and evolution.

\end{abstract}


\end{frontmatter}


\section{Introduction}

A circumstellar disk or ring is a relatively planar structure containing particulate matter and orbiting a star.  Disks are distinct from spherical 
clouds or envelopes of dust (and gas) that typically surround protostellar objects and some giant stars.  Circumstellar disks appear at every 
stage of stellar evolution, though the origin of the orbiting material is not always clearly understood.  Pre-main sequence stars accrete material 
from a disk that is the flattened remnant of the cloud from which they formed.  Disks found at young stars in subsequent evolutionary stages 
are the likely sites of planet formation, migration, and sometimes destruction.  Mature main-sequence stars exhibit dusty disks owing to recent 
or ongoing energetic collisions among orbiting asteroid or comet analogs -- sometimes referred to as the ``Vega phenomenon'' after the first
star observed to have orbiting, non-stellar material.  Several first-ascent and asymptotic giant stars are also known to have circumstellar 
disks, though their origins are still debated with hypotheses ranging from debris in a cold cometary cloud to consumed stellar companions.
It can be useful to consider debris disks as the entire planetary system, minus the planets; for detailed observational and theoretical reviews 
of debris disk phenomena, see \citet{zuc01} and \citet{wya08}.

The Solar System has a debris disk, part of which can be observed as zodiacal light, and has two major components in the asteroid 
and Kuiper belts, which provide a basic model for extrasolar debris disks.  Beginning with the discovery of circumstellar dust at Vega 
\citep{aum84}, it has gradually emerged that the majority of dusty main-sequence stars have (enhanced) disks analogous to the relatively icy 
and cold Kuiper belt \citep{wya03,bry06}.  In contrast to the zodiacal cloud of the Solar System, detectable extrasolar debris disks tend to be 
relatively clear of dust in their inner regions, with a few notable exceptions where warm analogs to the main asteroid belt have been detected 
\citep{bac09,su13}, and hot, transient dust likely due to recent giant impacts in the terrestrial zone \citep{son05,mel10b,men14}.  

Disks orbiting white dwarfs are a recently recognized phenomenon, with nearly 40 identifications since 2005, where the bulk required discovery 
and characterization from space.  These evolved systems are a sensitive probe of the frequency and long-term dynamics of planetary systems 
formed around intermediate-mass stars, and provide a unique window on planetesimal chemistry.  A journal review of this field is compelling as 
none exist outside of book chapters, and there is now more than a decade of infrared and optical observations of disks, including sensitive 
space-based data provided by the {\em Spitzer Space Telescope}.  Published theoretical studies of disk formation and evolution have increased 
in the last several years, and an overall comparison of theory with observation is needed.  

It is noteworthy that the known circumstellar disks orbiting white dwarfs have no analog among main-sequence or giant stars.  However, as
described in this review, it is likely that white dwarf planetary systems resemble, at least in part, both solar and extrasolar planetary systems,
including their planetesimal belts, while the disks themselves most resemble the planetary rings of the Solar System.  This article focuses on
the debris disk phenomenon for white dwarfs from a primarily observational perspective, but also discusses the relevant theoretical models; 
for an in-depth review of post-main sequence, planetary system evolution theory, see \citep{ver16}.  In the context of planetary system assembly 
and origins, it is fundamental that scientists can (indirectly) obtain extrasolar planetesimal compositions from the heavy element abundances of 
debris raining onto white dwarf atmospheres; for a review of this topic, see \citet{jur14}.  Partly in parallel, the aim of this article is to review the 
physical data and theory (from the stellar surface outwards) that makes the above scientific analysis possible, and to provide additional and 
broader context for white dwarf planetary systems.

The review begins in \S2 with a historical perspective, placing white dwarfs into the context of debris disk observations and introducing the 
pollution phenomenon.  Next, \S3 expounds on the singular influence of {\em Spitzer} and the rapid gain of observational knowledge during this 
era.  \S4 discusses the standard accretion processes and diagnostics, summarizes the overall observed properties of known polluted white dwarfs, 
and reviews the various surveys that have provided the best statistical inferences.  Some possible physical origins and evolution of the disks 
themselves are discussed in \S5, and a summary is provided in \S6.

\section{History and Background}

Owing to the bright and variable infrared sky, astronomical observations at these wavelengths where dust emission is most prominent are 
challenging; the historical detection of a debris disk orbiting the 5$^{\rm th}$ brightest star required the {\em Infrared Astronomical Satellite 
(IRAS)}.  Furthermore, because any (heated) circumstellar material derives its luminosity from the central star, the small radii of white dwarfs
implies orbiting dust can be remarkably hard to identify, as they are typically $\sim10^4$ times less luminous than their main-sequence 
counterparts.  Despite the unprecedented sensitivity and all-sky coverage of {\em IRAS}, not a single white dwarf was detected at any 
wavelength \citep{shi86}.  This is perhaps unsurprising given that {\em IRAS} was only sensitive to point sources brighter than 500\,mJy at 
its shortest wavelength bandpass of 12$\mu$m, while the brightest white dwarf in the sky (Sirius B) emits around 5\,mJy there \citep{ske11}.  

Nevertheless, the first dusty white dwarf was discovered within four years of the Vega discovery.  In strange contrast, however, white 
dwarfs are a relatively recent addition to the classes of stellar objects known to host circumstellar disks because additional examples 
remained hidden until 2005.  It is noteworthy that if white dwarfs only hosted debris analogous to main-sequence stars, their disks would 
still remain largely undetected today, even in the presence of dedicated and relatively advanced, infrared space missions.  Circumstellar 
dust at white dwarfs has not only been historically hard to detect, it remains considerably difficult or impossible to study in detail.  Only 
future space-based, and innovative ground-based, facilities can significantly impact the status quo for white dwarfs in the infrared.

Of particular relevance for white dwarfs is that infrared excess emission associated with a star can arise from irradiated circumstellar dust 
or from a self-luminous companion.  Owing to their compact nature, white dwarfs can be easily outshone in the near-infrared by low mass 
stellar and even brown dwarf companions.  Recognizing this fact, R. Probst searched a large sample of nearby white dwarfs for photometric 
excess at $1-2$\,$\mu$m to study the luminosity function of the lowest mass stars and brown dwarfs, as companions \citep{pro81,pro83}.  
This was the first search for infrared excess at white dwarfs, and Probst deserves the credit for an insight that is now taken for granted, and 
which fostered an abundance of subsequent infrared work on white dwarfs, including the first discovery of circumstellar dust.\footnote{For 
historical completeness and accuracy, \citet{pro83} detected $K$-band emission from the disk orbiting G29-38 sometime between 1978 
and 1980.  Although the single bandpass data were insufficient to confidently identify the excess emission, he gathered {\em the first 
photons from extrasolar planetary material(!)}, several years before the Vega and $\beta$ Pictoris disks were detected by {\em IRAS}.} 



\begin{figure}
\centering
\includegraphics[width=0.7\textwidth]{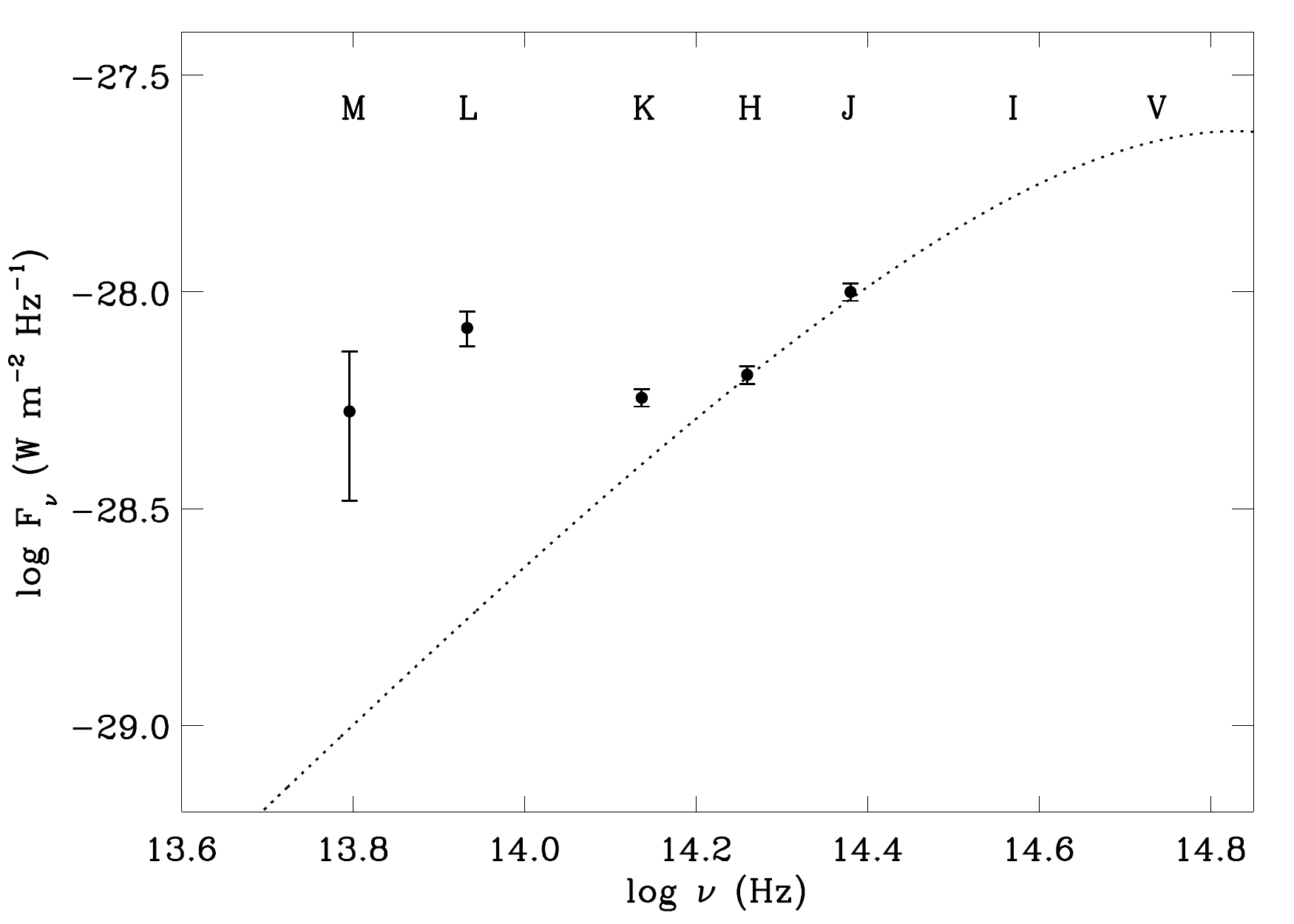}
\includegraphics[width=0.7\textwidth]{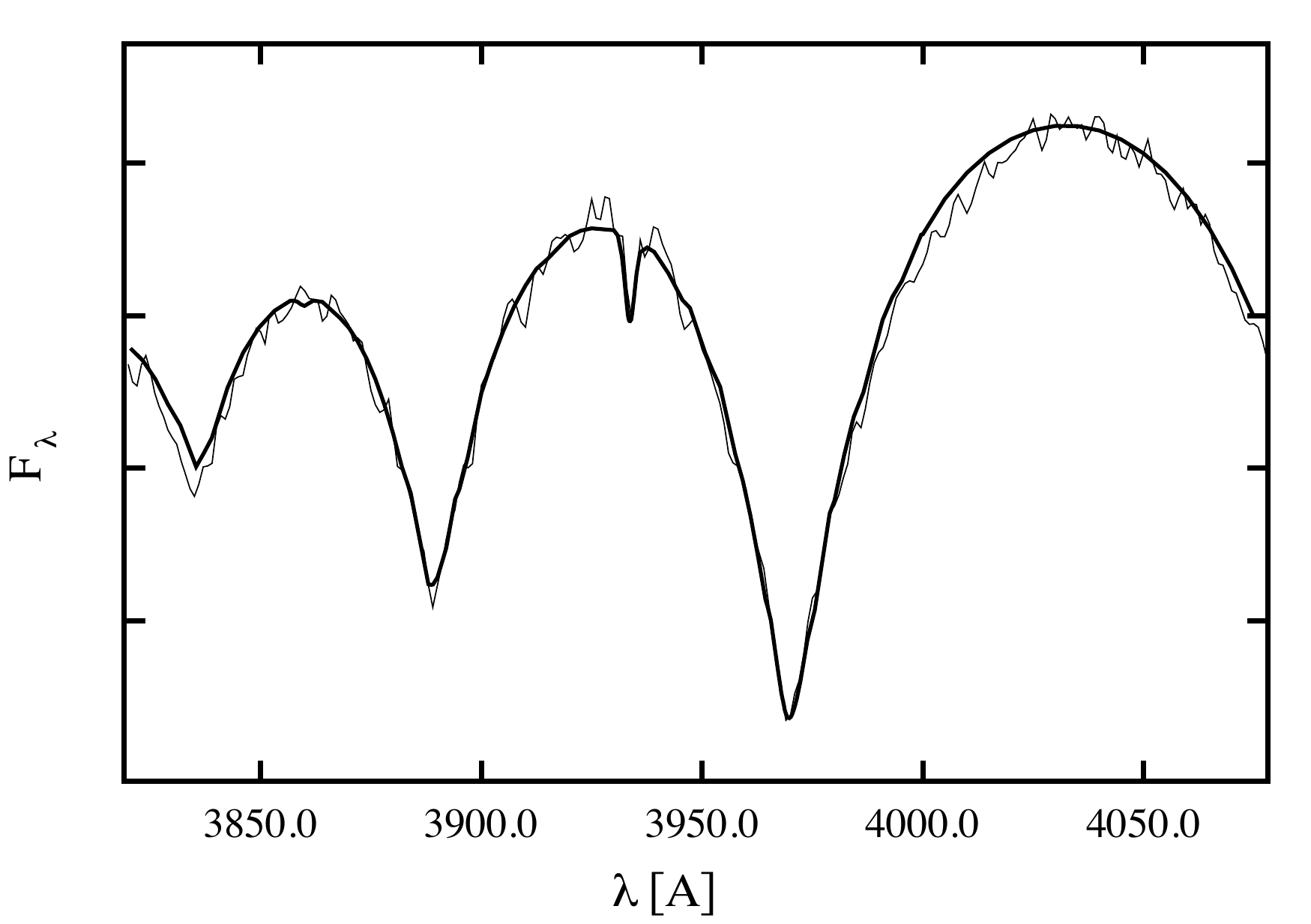}
\vskip -5 pt
\caption{The prototype dusty white dwarf G29-38.  {\em Upper panel}: The discovery of infrared excess made at the IRTF.  The plot is 
a reproduction of the figure presented in \citet{zuc87} and shows their measurements with errors in five infrared bandpasses labelled in 
the figure $J$ (1.25\,$\mu$m), $H$ (1.65\,$\mu$m), $K$ (2.2\,$\mu$m), $L$ (3.5\,$\mu$m), and $M$ (4.8\,$\mu$m).  The dotted line is 
a blackbody consistent with the $T_{\rm eff}=11\,500$\,K stellar photosphere in the infrared.  {\em Lower panel}: The detection of 
photospheric calcium in the optical \citep{koe97}, at the same atomic transition that is strongest in the Sun (the Fraunhofer K line).  
The thin line is the actual data while the thick line is a model fit.  The Ca\,{\sc ii} absorption appears rather weakly beside the 
pressure-broadened Balmer lines in this hydrogen-rich star.  
\label{fig1}}
\end{figure}

\subsection{The Discovery of Infrared Excess at G29-38}

Inspired by the work of Probst and armed with a rapidly evolving set of detectors at the NASA Infrared Telescope Facility (IRTF), 
\citet{zuc87} identified the first white dwarf with infrared excess emission that was not associated with a stellar companion; G29-38.  
Photometric observations at three bandpasses longward of 2\,$\mu$m revealed $T\approx1200$\,K emission in excess of that expected 
for the stellar photosphere of this relatively cool white dwarf (Figure \ref{fig1}).  

However, as the observations were driven by a search for substellar companions, the infrared emission was initially attributed to a 
spatially unresolved brown dwarf.  Heated circumstellar dust was not favored due to the likelihood of rapid dissipation by radiation drag, 
with \citet{zuc87} prophetically remarking that if a debris disk were orbiting sufficiently close to attain such high temperatures, then spectral 
signatures of accretion should be seen.

Over the next few years, the infrared emission at G29-38 was studied intensely by many groups, and its unique properties sparked 
interest across several subfields of astrophysics research: brown dwarf and planet hunters, asteroseismologists, infrared astronomers, and 
white dwarf pundits.  Observational evidence gradually began to disfavor a brown dwarf as the source of infrared emission.  First, some of 
the first near-infrared imaging arrays revealed G29-38 to be a point source in several bandpasses.  Second, near-infrared spectroscopy
measured a continuum flux source \citep{tok88}, whereas a very cool atmosphere was expected to exhibit absorption features.  Third, 
the detection of optical stellar pulsations echoed in the near-infrared were difficult to reconcile with a brown dwarf secondary \citep{pat91,
gra90}.  Fourth, significant 10\,$\mu$m emission was detected at G29-38, a few times greater than expected for a cool object with the 
radius of Jupiter, essentially ruling out the brown dwarf companion hypothesis \citep{tok90,tel90}.

A decade after the discovery of its infrared excess, the optical and ultraviolet spectroscopic detection of multiple metal species in the 
atmosphere of G29-38 \citep{koe97} made it clear the star is currently accreting from its circumstellar environs (Figure \ref{fig1}).

\subsection{White Dwarf Atmospheric Pollution}

It would not be possible to accurately review circumstellar disks at white dwarfs, including the pivotal role of G29-38, without introducing 
the phenomenon of atmospheric metal contamination.  Rather than being a central focus of this review, the topic of photospheric metals will 
be discussed in the context of, and its relation to circumstellar material around white dwarfs.

The origin and abundances of photospheric metals in isolated white dwarfs has been an astrophysical curiosity dating back to the era when 
the first few white dwarfs were finally understood to be subluminous via the combination of spectra and parallax \citep{van19}.  In a half page 
journal entry, \citet{van17} reported (via W. Adams) that his accidentally discovered faint star with large proper motion had a spectral type of 
``about F0'', almost certainly based on its strong Ca\,{\sc ii} H and K absorption features (Figure \ref{fig2}).\footnote{With hindsight, the 
detection of metals in vMa\,2 in 1917 was the first observational evidence of planetary systems beyond the Sun \citep{zuc15}.}

Any primordial heavy elements in white dwarfs can only be sustained in their photospheres for the brief period while the star is still 
rather hot and contracting, and then only to a certain degree \citep{cha95,bar14b}.  For $T_{\rm eff}<25,000$ K, gravitational settling is 
enhanced by the onset of convection and atmospheric heavy elements sink rapidly in the high surface gravities of white dwarfs \citep{fon79,
vau79} leaving behind only hydrogen or helium.  Downward diffusion timescales for heavy elements in cool white dwarfs are always orders 
of magnitude shorter than their evolutionary (cooling) timescales \citep{paq86}, and thus external sources must be responsible for any 
photospheric metals.

\subsection{Stellar, Interstellar, or Circumstellar Matter}

There are only three possible sources for the atmospheric metals seen in isolated white dwarfs: 1) stellar material either primordial or
fallback, 2) accretion from the interstellar medium, or 3) infall from an intrinsic circumstellar environment (i.e.\ a remnant planetary system).  
Regardless of origin, the deposition of heavy elements necessary to enrich a white dwarf atmosphere may be facilitated by the formation of
a disk, and thus photospheric metals and circumstellar disks are likely to have a fundamental physical connection.


\begin{figure}[h!]
\centering
\includegraphics[width=0.7\textwidth]{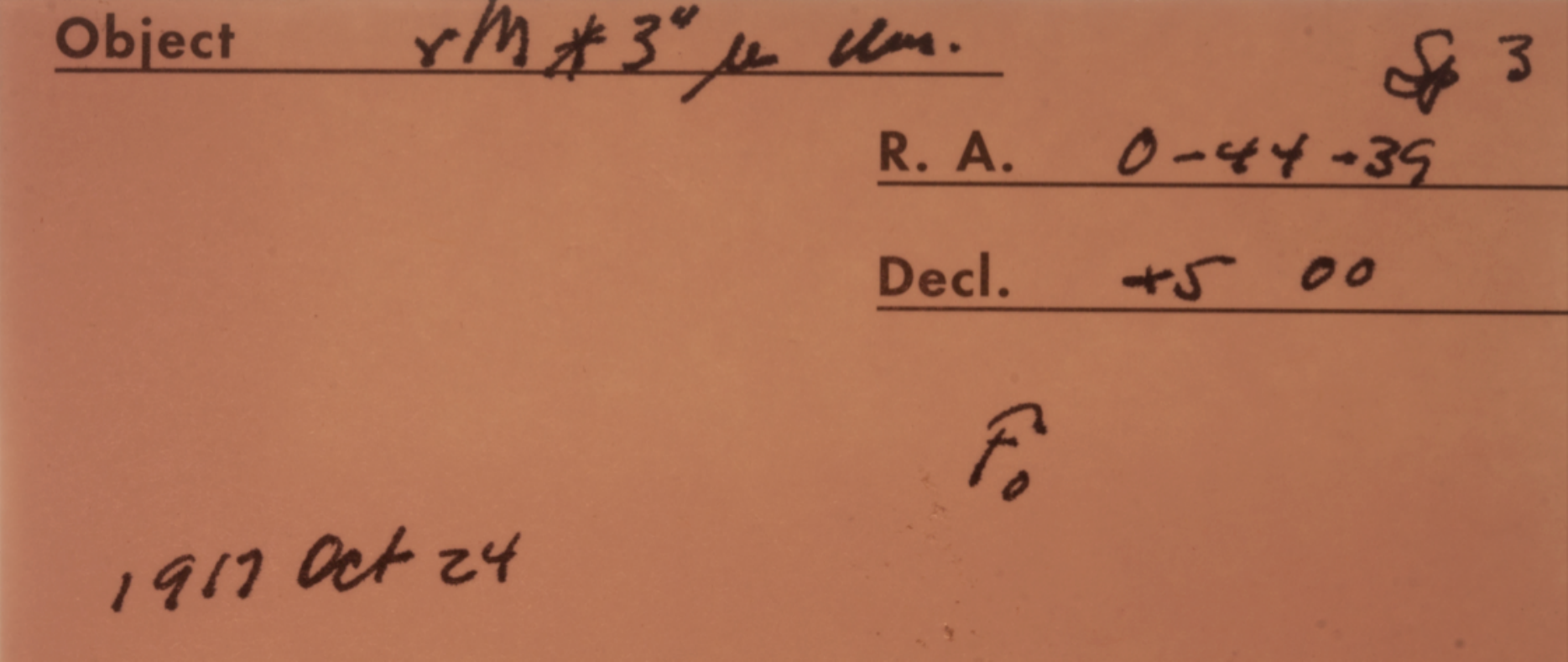}
\vskip 5 pt
\includegraphics[width=0.7\textwidth]{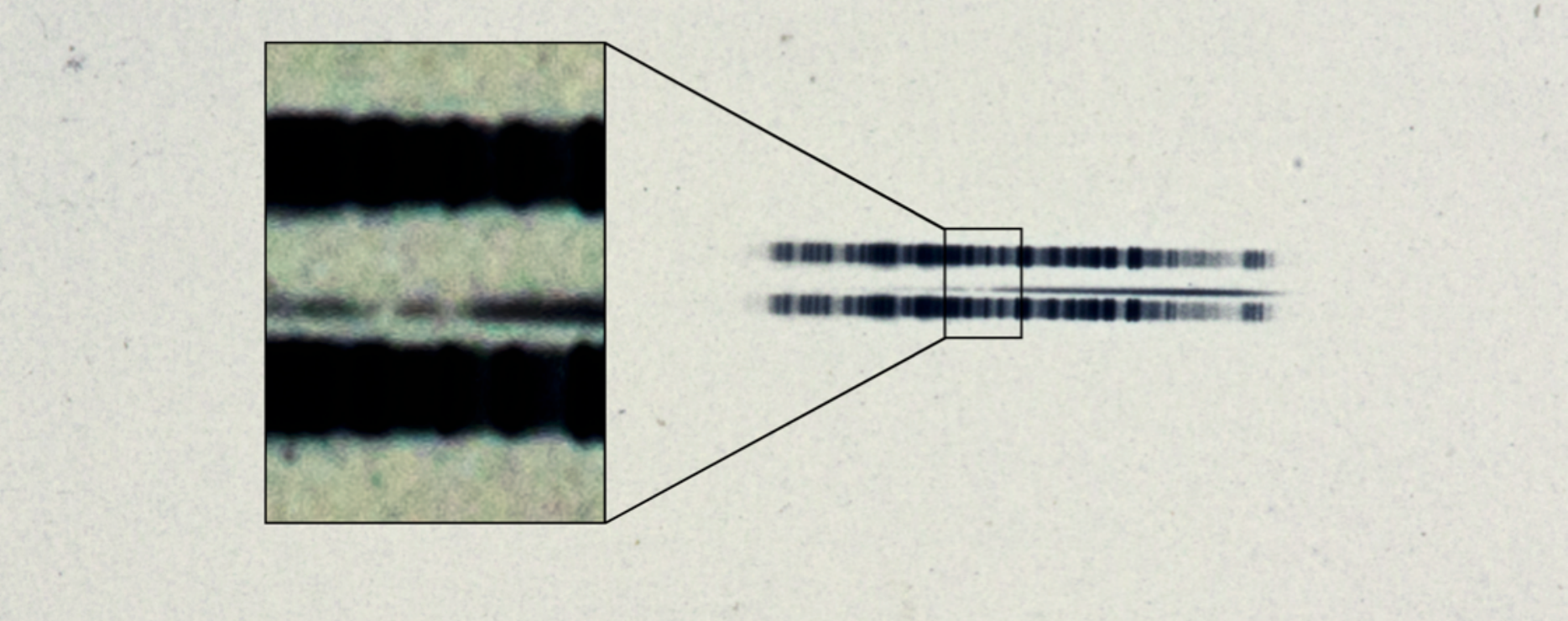}
\vskip 0 pt
\includegraphics[width=0.7\textwidth]{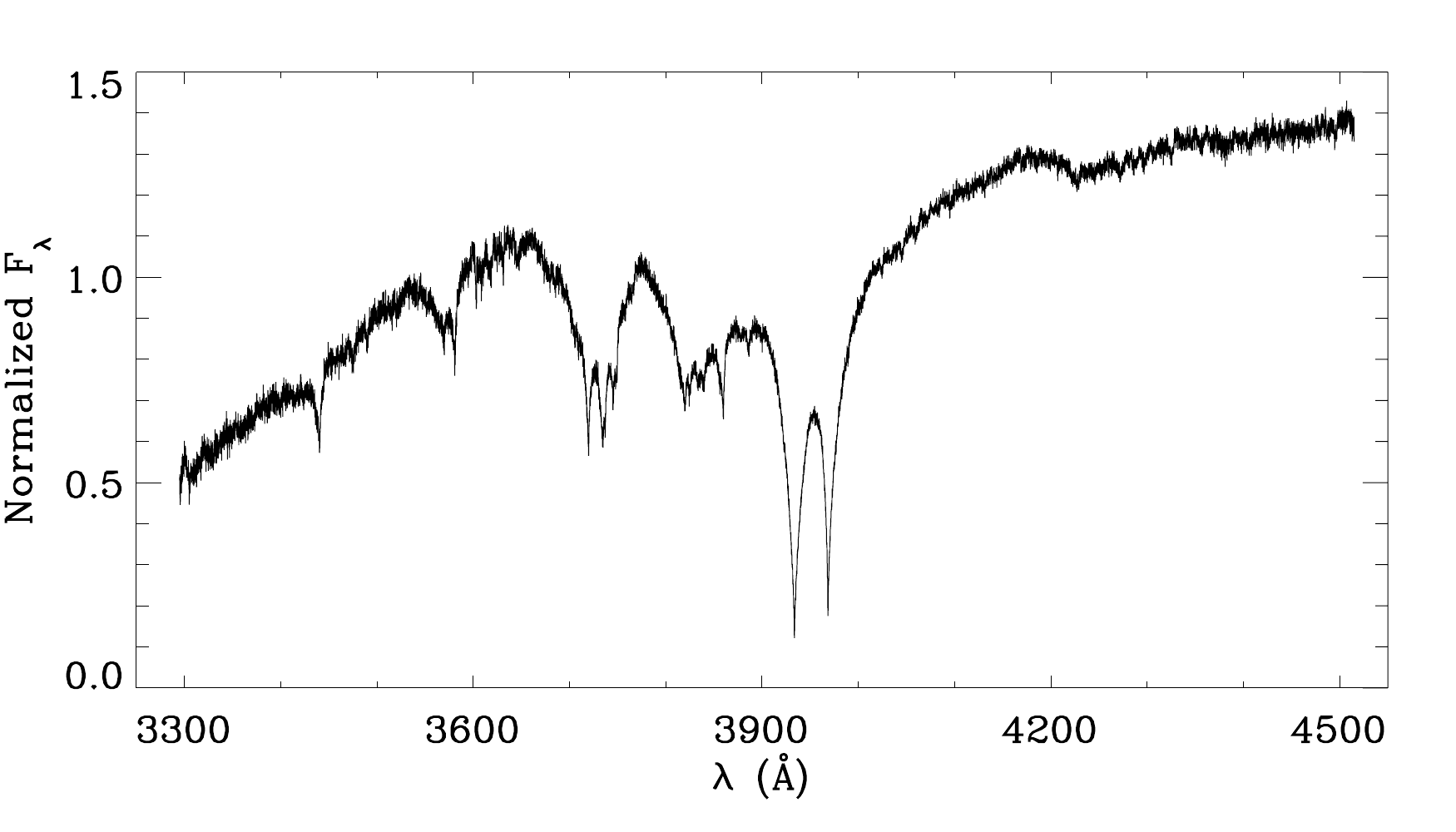}
\vskip -10 pt
\caption{The prototype metal-enriched white dwarf vMa\,2.  {\em Upper panels}:  The 1917 photographic plate spectrum of vMa\,2, and plate 
sleeve with handwritten notes by observer W. S. Adams \citep{van17}; these historical records were kindly supplied by Carnegie Observatories, 
who maintains the Mt.\ Wilson archives (J. S. Mulchaey 2015, private communication).  Box highlighted are the strong Ca\,{\sc ii} H and K 
Fraunhofer lines, and are relatively easy to see in the century old spectrum.  In contrast to hydrogen-rich atmospheres (e.g.\ Figure \ref{fig1}), 
metal absorption features in helium-rich stars can be prominent, and often dominate their optical spectra.  {\em Lower panel}:  An unpublished 
optical spectrum of vMa\,2 taken with UVES on the Very Large Telescope (VLT).  All salient features are absorption due to Fe, Mg, or Ca.
\label{fig2}}
\end{figure}

The argument against a primordial origin for atmospheric heavy elements is the fact that they rapidly sink in the high gravity 
environment of white dwarfs.  This theoretical expectation is corroborated by high-resolution spectra that typically exhibit pure hydrogen or 
helium compositions \citep{zuc03,koe05}.  Once the stellar ember has cooled below 25\,000\,K, metals can no longer be radiatively supported 
and diffuse from the atmosphere in a matter of days to years \citep{koe09}.  At least 20\,Myr is required to reach this temperature, and roughly 
0.5\,Myr is spent as a hot and luminous star with $T_{\rm eff}\sim10^5$\,K, $L\sim50$\,$L_{\odot}$; \citep{fon01}.  During this time, any 
circumstellar gas and small particles ($a<1$\,cm) within several AU, such as dust formed within a stellar outflow during the giant branches, 
will be effectively removed from the system by engulfment, sublimation, radiation pressure, or drag forces.  Because white dwarf disks and 
pollution are observed at $30-600$\,Myr cooling ages, the circumstellar material must derive from relatively parent large bodies, or originate 
beyond several AU.

Historically, accretion from the interstellar medium was the most widely accepted hypothesis for metals observed in white dwarfs.  This 
interpretation was at least partly due to the fact that until 1997, all but one case of photospheric metals had been detected in relatively 
cool, helium-rich stars (spectral type DZ or DBZ; \citealt{sio90b}).  Such white dwarfs have relatively transparent outer layers that aid the 
detection of heavy elements (compare e.g.\ Figures \ref{fig1} and \ref{fig2}), and relatively deep convection zones with metal diffusion 
timescales up to $10^6$\,yr \citep{koe09}.  The relatively long, atmospheric residence times in these stars allow for the possibility that 
their metals could be the residuals of interstellar cloud encounters \citep{dup93a,dup93b}.  

The interstellar accretion model faced two major obstacles even before dust disks emerged as a class of objects associated with
metal-rich white dwarfs.  Fundamentally, the nearly universal lack of hydrogen in DBZ stars has been a glaring problem, even for the
larger class of helium atmosphere white dwarfs \citep{koe76,wes79,aan93}.  In essence, if white dwarfs accrete sufficient material of 
roughly solar composition, they should all (eventually) exhibit hydrogen-dominated outer layers, whereas a significant fraction of all cool 
white dwarfs are hydrogen-deficient stars, including many with metals.  Another serious challenge came from the detection of metals in 
stars with hydrogen-rich atmospheres (spectral type DAZ), where the heavy element diffusion timescales are only days to years.  The 
first was confirmed in 1983 \citep{lac83}, but remained anomalous until 1997 when two additional examples were found, including G29-38 
\citep{hol97,koe97}.  A targeted search brought further detections the following year \citep{zuc98}, and because these stars must have 
experienced recent or ongoing accretion, the DAZ white dwarfs implied stringent constraints for interstellar accretion scenarios.

\clearpage

These observational developments and outstanding theoretical issues led some to consider circumstellar accretion models.  The earliest
of these was the cometary impact model put forth by C. Alcock, and represented the first recognition of white dwarf pollution as a potential 
exoplanetary system signature \citep{alc86}.  A clear strength of cometary impacts is greater consistency with the hydrogen-poor nature 
of the accreted material in numerous DBZ stars, although it was soon noted that volatile-depleted planetary material was more appropriate 
\citep{sio90a}.  Other difficulties for the comet cloud model came from two DAZ stars; G238-44 has a sufficiently short metal diffusion time 
that requires a continuous rain of comets \citep{hol97}, while the circumstellar debris around G29-38 could not be generated by an impact 
\citep{zuc98}.

\subsection{The Asteroid Disruption Model}

Two major developments occurred in 2003, representing a historic crossroads for white dwarf disks and pollution just prior to the launch 
of the {\em Spitzer Space Telescope} \citep{wer04}.  \citet{zuc03} published a high-resolution spectroscopic study of photospheric metals 
in a large sample of cool, hydrogen-rich white dwarfs, finding several new examples of DAZ stars and establishing a minimum frequency 
of 20\% for the pollution phenomenon in single stars.  The results were shown to be inconsistent with the cometary impact model in the
following way: it was predicted that there should be sharp rise in the number of DAZ with decreasing calcium abundance, owing to more 
frequent impacts of smaller objects \citep{alc86}, but instead a relatively flat distribution was found in calcium abundance over several 
orders of magnitude.  Neither could the DAZ sample be reconciled with accretion from interstellar medium, due to the lack of dense clouds 
in the Local Bubble, a result also found previously for a modest number of DBZ stars \citep{aan93}.

With the sole exception of G29-38 there was a distinct lack of reliable (infrared) data on the circumstellar environments of white dwarfs, 
yet a growing profusion of stars contaminated by metals, and problems with existing hypotheses.  In a short but seminal paper, \citet{jur03} 
modeled the observed properties of G29-38 by invoking a tidally-destroyed minor planet (i.e.\ asteroid) that generates an opaque, flat disk
of dust analogous to the rings of Saturn.  Rather than impacting the star, an asteroid perturbed into a highly eccentric orbit passes within 
the stellar Roche limit and is torn apart by gravitational tides \citep{deb02}.  Ensuing collisions reduce the fragments to rubble and dust, 
dynamically relaxing into a flat configuration over multiple orbital timescales.  The closely orbiting dust is heated by the star producing an 
infrared excess, and slowly rains down onto the stellar surface, polluting its otherwise-pristine atmosphere with heavy elements.  

The tidally disrupted asteroid model, and accretion from a remnant exoplanetary system, is now the standard framework for white dwarf
metal pollution.  Interestingly and years prior, \citet{gra90} deduced that the closely orbiting dust at G29-38 must have resulted from a 
relatively recent, catastrophic event near the white dwarf, such as a near collision with an asteroid or comet.  This qualitative model for 
the infrared emission at G29-38 invoked cm-sized pebbles as these would not be rapidly destroyed near the star, but the flat disk model
achieves this by shielding the bulk of material from the full stellar radiation field.  In the sections that follow, it will be shown that this model
has been strongly corroborated over the past decade by myriad infrared, optical, and ultraviolet data.

\section{The {\em Spitzer} Revolution and Era}

Fortunately, the last decade has seen a profusion of white dwarf disk discoveries, due primarily to the 2003 launch and unprecedented
infrared performance of {\em Spitzer}.  Important roles have also been played by projects with large sky coverage such as the Sloan Digital 
Sky Survey (SDSS; \citealt{gun06}), the Two Micron All Sky Survey (2MASS; \citealt{skr06}), and the {\em Wide Field Infrared Explorer (WISE}; 
\citealt{wri10}).  The number of confirmed metal-rich white dwarfs with circumstellar dust consisted of a single system from 1987 until 2005, 
but has grown to nearly three dozen (and counting) within a decade.  

Although there have been some infrared disk detections from the ground longward of 2.5\,$\mu$m \citep{tok90,bec05,mel11}, most systems 
are (far) too faint to be studied in this manner, and all confirmed dusty white dwarfs at the time of writing have been observed with {\em Spitzer}.  
As demonstrated in this section, it is only beyond 2.5\,$\mu$m that the bulk of infrared excesses become salient, and thus the final NASA Great 
Observatory played a leading and fundamental role in this field, and specifically in connecting atmospheric metals in white dwarfs to circumstellar 
debris.

\subsection{Basic Dust Properties}

\begin{figure}[h!]
\centering
\includegraphics[width=1.0\textwidth]{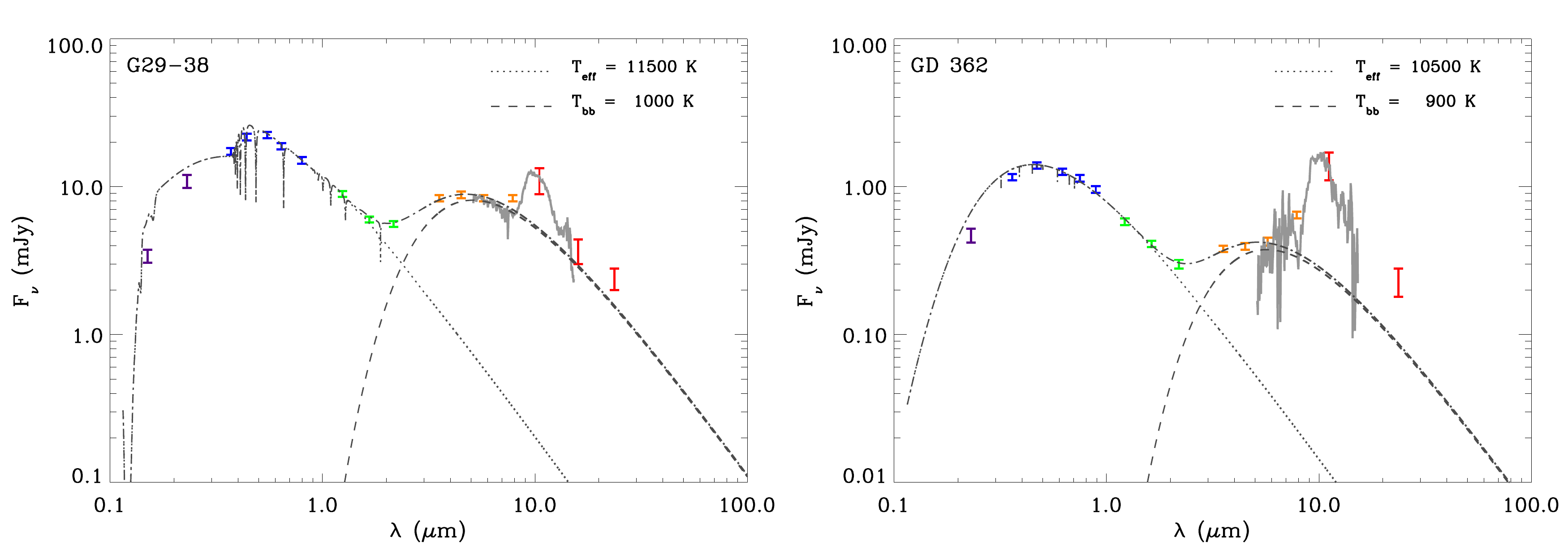}
\vskip -10 pt
\caption{The first two dusty and metal-enriched white dwarfs observed with all three cryogenic instruments aboard {\em Spitzer} 
\citep{rea05a,jur07b}.  Colored error bars represent photometric data: {\em GALEX} far- and near-ultraviolet (purple); optical $UBVRI$ or 
$ugriz$ (blue); near-infrared $JHK$ (green); IRAC $3-8\,\mu$m (orange); $N$-band, IRS 16\,$\mu$m, and MIPS 24\,$\mu$m (red). IRS 
$5-15\,\mu$m spectra are also shown (solid grey), together with stellar atmospheric models (dotted grey), and blackbody approximations 
to the disk emission (dashed grey).  In contrast to the bulk of debris disks at main-sequence stars, there is a notable lack of dust 
significantly cooler than 1000\,K.
\label{fig3}}
\end{figure}

Within the first two cycles of {\em Spitzer} observations, the only two known white dwarfs with circumstellar dust (G29-38 and 
GD\,362\footnote{Ironically, the first dusty white dwarf to be discovered in the {\em Spitzer} era, and the second known example of the 
phenomenon, was identified from the ground \citep{bec05,kil05}.}) were observed with all three cryogenic instruments \citep{rea05a,jur07b}.  
The Infrared Array Camera (IRAC; \citealt{faz04}) obtained photometry at 3.6, 4.5, 5.7, and 7.9\,$\mu$m, the Infrared Spectrograph (IRS; 
\citealt{hou04}) acquired $R\sim100$ spectra between 5 and 15\,$\mu$m, and the Multi-band Imaging Photometer for {\em Spitzer} (MIPS; 
\citealt{rie04}) measured fluxes at 24\,$\mu$m for the two earliest examples shown in Figure \ref{fig3}, and later for a number of similar 
white dwarf targets prior to the loss of cryogen in 2009.


\begin{figure}[h!]
\centering
\includegraphics[width=1.0\textwidth]{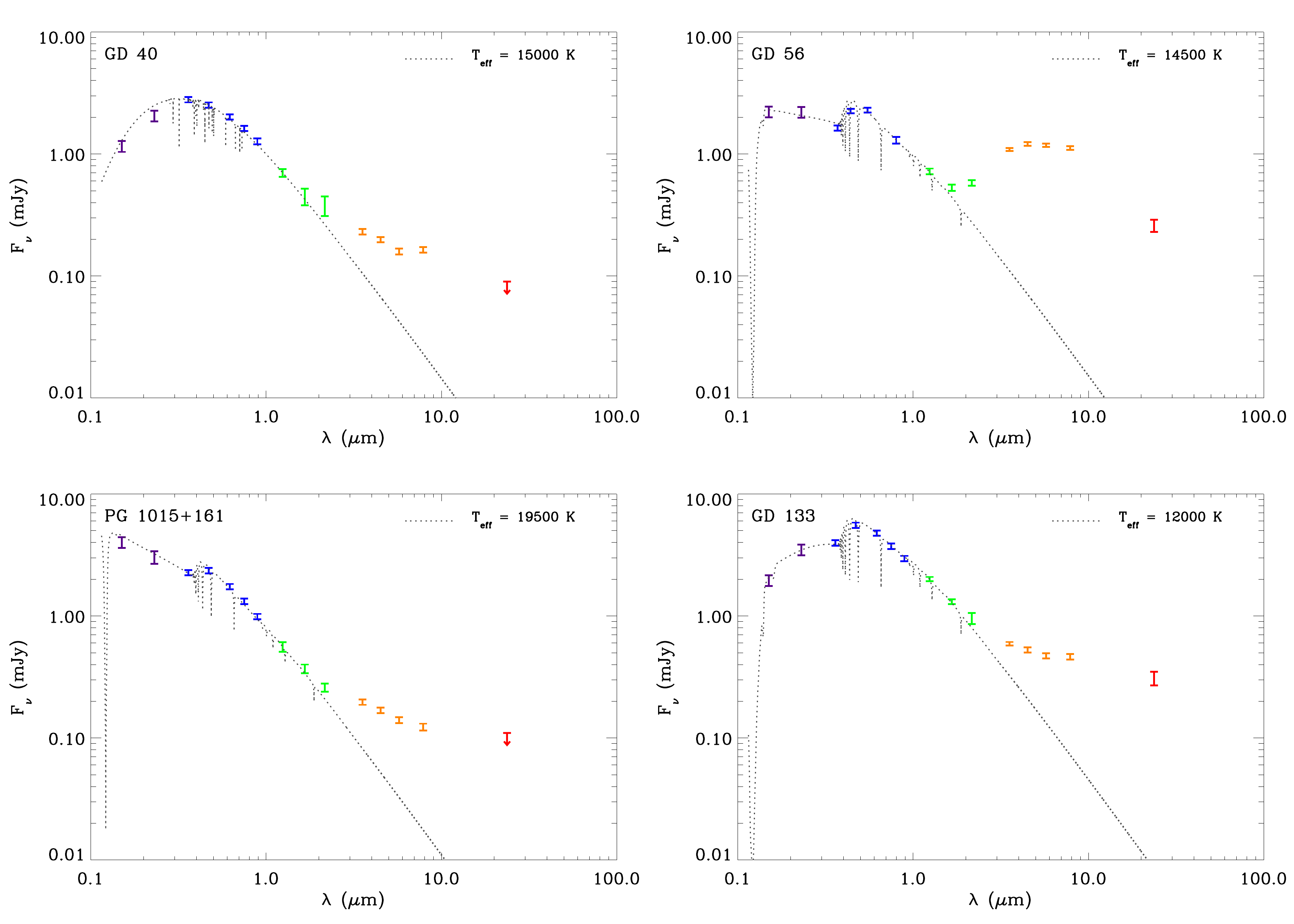}
\vskip -10 pt
\caption{{\em Spitzer} IRAC and MIPS photometry of four polluted white dwarfs with circumstellar dust \citep{jur07b}, observed within the 
first two General Observer cycles (symbols as in Figure \ref{fig3}).  As with the earlier cases, the infrared excesses begin at $2-3$\,$\mu$m 
and share a striking lack of emission from cool dust, as emphasized by the upper limits at 24\,$\mu$m for two targets (downward arrows).  
\label{fig4}}
\end{figure}

There are a few remarkable, shared properties that can be seen in these two infrared datasets, two of which turn out to be ubiquitous 
(to date) among dusty and polluted white dwarfs.  First are the strong silicate emission features \citep{rea05a,jur07b}.  In fact, the emission 
observed at GD\,362 is among the strongest silicate features seen towards any mature star (compare e.g.\ \citealt{men15,mel10b}).  Second, 
there is a notable lack of relatively cool dust, unlike the bulk of all debris disks orbiting main-sequence stars that exhibit infrared excess only 
at 24 or 70\,$\mu$m MIPS observations \citep{su06,wya08}.  While models indicate dusty stars such as Vega harbor substantial planetary 
material consistent with orbital regions analogous to the Kuiper Belt, those same (blackbody) models would place 1000\,K dust orbiting these 
two white dwarfs within 0.7\,$R_{\odot}$ (roughly 50 stellar radii).  The next wave of disks discovered and characterized photometrically by 
{\em Spitzer} strongly suggested that a lack of cool material was a hallmark of dust orbiting polluted white dwarfs, as consistently indicated 
either by low (or undetected) 24\,$\mu$m MIPS fluxes, relative to those at IRAC wavelengths (Figure \ref{fig4}).  

A third shared property among these first half dozen examples is that the infrared excesses all become apparent by $2-3\,\mu$m, a trait 
that is now known to be common (but not universal), and hence the most prominent of these are detectable from the ground using $K$-band 
spectroscopy \citep{kil06,kil07}.  Although these first several discoveries are likely biased by favoring the bright end of the disk luminosity 
function, they are highly instructive.  The emergence of dust emission over the stellar photosphere at these wavelengths indicates there 
is substantial material with $T=(2898$\,K\,$\mu$m$^{-1} / \lambda) \approx1000-1500$\,K, demonstrating the circumstellar debris orbits 
sufficiently close to pollute the stellar surface.  Thus, these findings encouraged continued searches for thermal dust emission around 
metal-enriched white dwarfs \citep{jur07a}.


\subsection{Theoretical Considerations}

The first several disk discoveries were accompanied by modest theoretical work, but fundamental in the context of {\em Spitzer} data, 
the asteroid disruption model, and the ubiquitous atmospheric metals.  Optically thin circumstellar dust cannot long survive such close 
orbits as those inferred for blackbody grains such as modeled in Figure \ref{fig3}.  The luminosities of typical dusty white dwarfs ($10^{-3}
L_{\odot}\lesssim L\lesssim10^{-1.5}L_{\odot}$) are insufficient to remove grains of any size by radiation pressure \citep{far08a}, and thus 
in stark contrast to debris orbiting main-sequence stars, there is no lower limit to grain sizes in white dwarf disks.  Despite this fact, 
Poynting-Robertson (PR) drag will efficiently remove any closely-orbiting, optically thin, small dust grains around these stars \citep{han06}.  
The timescale for particles of radius $a$ and density $\rho$ to spiral inward from orbital radius $r$ towards a star of luminosity $L$ is 
\citep{bur79}

\begin{equation}
t_{\rm pr} =  \frac{4\pi}{3} \ \frac{c^2 r^2 \rho a}{LQ_{\rm pr}}
\label{eqn1}
\end{equation}

\medskip
\noindent
Here, $Q_{\rm pr}$ is the coupling coefficient for radiation pressure and is $\approx1$ for the case of geometric optics (appropriate for 
micron-sized and larger particles; \citealt{art88}).  For the luminosity range given above, PR drag will typically remove any unshielded, 
micron-sized and smaller, $T\sim1000$\,K dust in less than 10\,yr.

Hence small particles expected to be the most efficient emitters at {\em Spitzer} wavelengths must be replenished on short timescales 
or persist in an optically thick configuration.  While it is possible in principle to replenish an entire, optically thin disk of small particles on
decade timescales, because at least 20\% of white dwarfs are externally polluted over a wide range of metal sinking timescales (e.g.\ 
\citealt{zuc03}), this would imply an influx of star-grazing planetesimals, sustained over long periods, and is therefore unrealistic (as for 
the cometary impact model).  The tidally-disrupted asteroid model circumvents this problem by arranging the debris in a flat and (vertically) 
opaque configuration \citep{jur03}.  In this geometry, the disk height to width ratio is negligible, and only a tiny fraction of particles are subject 
to the full stellar radiation and drag force.  For stellar radius $R$ and effective temperature $T_{\rm eff}$, dust temperature at radius $r$ is 
then determined by \citep{chi97}

\begin{equation}
T \approx { \left( \frac{2}{3 \pi} \right) }^{1/4} { \left( \frac{R}{r} \right) }^{3/4} T_{\rm eff}
\label{eqn2}
\end{equation}

\medskip
\noindent
It is worth noting that particles in such a ring are cooler than exposed blackbody grains at the same orbital radius.

The flat ring model is fully characterized by stellar $R$ and $T_{\rm eff}$, the distance to the system $d$, and three free parameters: 
inner and outer disk radii $r_{\rm in}$ and $r_{\rm out}$ (fixing $T_{\rm in}$ and $T_{\rm out}$), and inclination $i$ to the line of sight .  
The flux from the ring is \citep{jur03}

\begin{equation}
F = \frac{2 \pi \cos{i}}{d^2} \int_{r_{\rm in}}^{r_{\rm out}} B_{\nu}(T) \ r \ dr
\label{eqn3}
\end{equation}

\medskip
\noindent
where $B_{\nu}(T)$ is the Planck function.  Substituting $x=h\nu/kT$ with dust temperature given as in Equation \ref{eqn2}, the emission 
from the flat disk can be calculated using

\begin{equation}
F \approx 12 \pi^{1/3} \ \frac{R^2 \cos{i}}{d^2} {\left( \frac{2kT_{\rm eff}}{3h\nu} \right)}^{8/3} \ \frac{h\nu^3}{c^2} \int_{x_{\rm in}}^{x_{\rm out}} \frac{x^{5/3}}{e^x-1} \ dx
\label{eqn4}
\end{equation}

\begin{figure}[h!]
\centering
\includegraphics[width=1.0\textwidth]{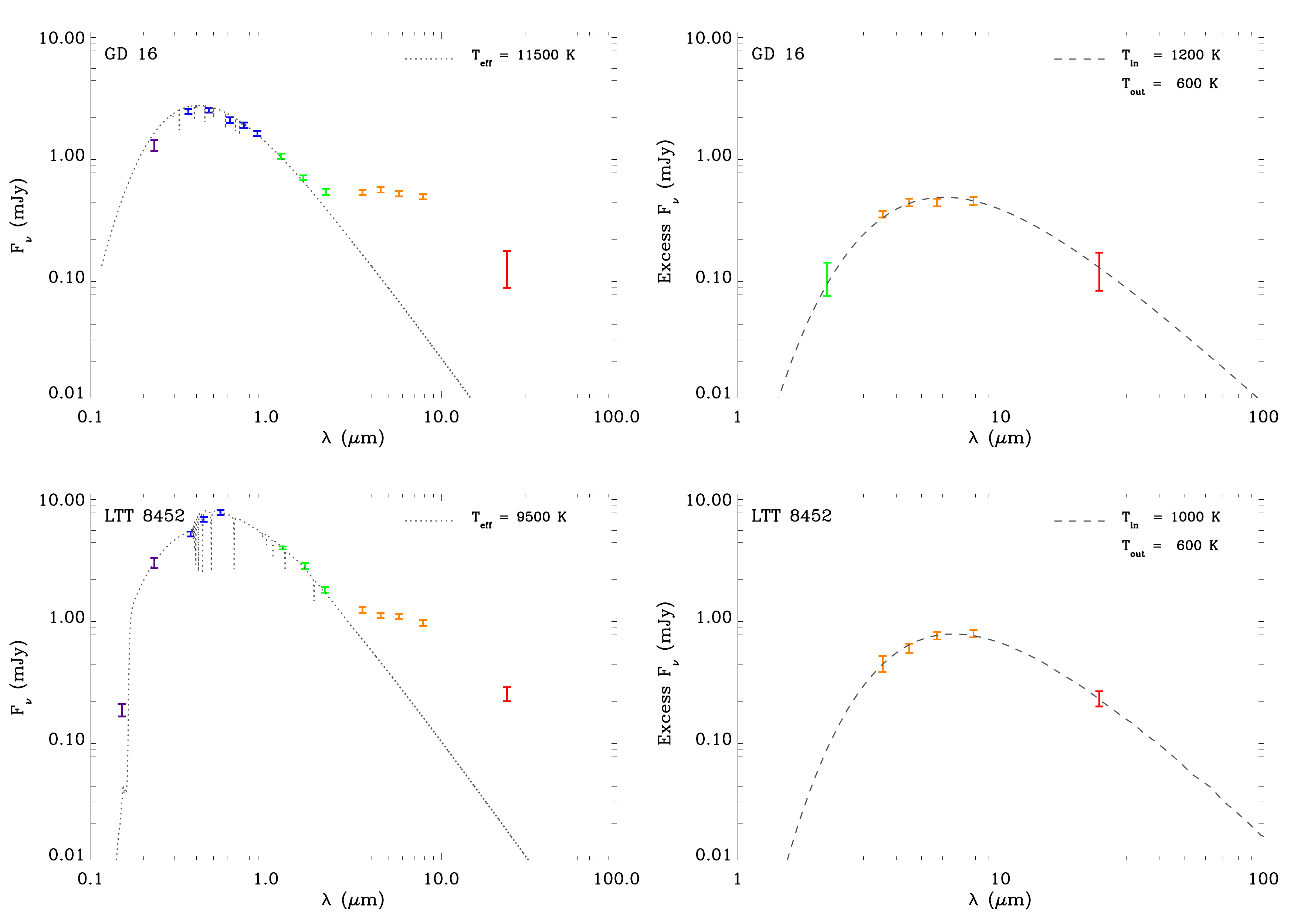}
\vskip -10 pt
\caption{Two examples of flat disk models fitted to {\em Spitzer} IRAC and MIPS flux measurements (symbols as in Figure \ref{fig3}).  The 
left hand panels show the full spectral energy distributions together with stellar atmospheric models \citep{von07,far09}, while the right hand 
panels show the excess fluxes after subtraction of the photospheres, and are overplotted with the disk models.  It is noteworthy that one of 
the stars does not exhibit an infrared excess in $K$-band photometry.  Both of these particular disks are well modeled with inclinations near 
50$^{\circ}$; the inferred dust temperatures at the ring inner and outer edges are given in each panel, and implying the disks are entirely 
contained within 1\,$R_{\odot}$.
\label{fig5}}
\end{figure}

\medskip
Figure \ref{fig5} displays examples of the flat ring model as applied to {\em Spitzer} IRAC and MIPS photometry for two polluted white 
dwarfs \citep{far09}.  As with Figures \ref{fig3} and \ref{fig4}, these two infrared excesses arise between 2 and 3\,$\mu$m, and exhibit the 
characteristic decreasing flux towards 24\,$\mu$m.  In both cases the disk models reproduce the measured flux excesses rather well, 
with inner disk temperatures similar to that expected for blackbody grains, but with modestly cooler extensions that better account for the 
8 and 24\,$\mu$m emissions.  These two stars provide a good showcase for the flat ring model, as their silicate emission features (see \S3.3 
and Figure \ref{fig6}) do not significantly alter their mid-infrared photometry (as for GD\,362 in Figure \ref{fig3}).  More sophisticated models 
are needed to simultaneously fit infrared spectral features and broad-band fluxes, but the flat ring model is the basis.


The flat disk models naturally possess a modest degree of uncertainty; $T_{\rm in}$, $T_{\rm out}$, and $i$ are somewhat degenerate in 
how they combine to determine the viewed disk solid angle.  Generally speaking, the inner disk edges are fairly well constrained by their 
2.2 and 3.6\,$\mu$m emission, with typical allowed ranges of 10\% in $T_{\rm in}$ (and hence inner disk radius), and up to 20\% at the 
extremes of what might be reasonably fitted \citep{jur07a}.  Longer wavelength disk fluxes, however, can often be reproduced by either 
a relatively wide temperature and radii range seen at high inclination, or a relatively narrow temperature and radii range seen at modest 
inclination \citep{gir12,ber14}.  Despite this degeneracy in the outer disk properties, a significant population of high inclination systems is 
statistically unlikely and hence modest inclinations and narrow radial extents are favored \citep{roc15}.

Although the morphological appearance of the Figure \ref{fig5} models can be approximated by single temperature blackbodies (as in
Figure \ref{fig3}) of $T\approx(T_{\rm in} + T_{\rm out})/2$, the flat ring has important physical context.  Excluding the 8\,$\mu$m bandpass 
that partly overlaps the region of silicate emission, excellent fits to the thermal continua of all but one of the stars in Figures \ref{fig3}--\ref{fig5}
have been achieved using flat ring models \citep{jur07a,far08a}.  For all these well-modeled systems, the resulting inner and outer disk radii lie
in the ranges $0.1-0.2\,R_{\odot}$ and $0.3-0.4\,R_{\odot}$, respectively.  These outer disk radii can be compared directly to the stellar Roche 
limit, inside of which tidal disruptions are expected.

The distance at which a lesser orbiting body of density $\rho$ is tidally split by a greater body of density $P$ can be expressed as 

\begin{equation}
\delta \approx \alpha { \left( \frac{P}{\rho} \right) }^{1/3} R
\label{eqn5}
\end{equation}

\medskip
\noindent
where $\alpha$ is a constant which typically has a value in the range $1-2.5$, depending on multiple factors \citep{dav99}.  The classical 
value of $\alpha=2.45$ is for the case of a rotating, uniform, self-gravitating liquid in a circular orbit; whereas in the rigid sphere approximation 
appropriate for stray asteroids the value becomes $\alpha=1.26$.  For asteroid densities in the range $\rho=1.0-3.5$\,g\,${\rm cm}^{-3}$ and a 
typical white dwarf of $0.6\,M_{\odot}$ and 0.013\,$R_{\odot}$, this results in $\delta\approx0.8-1.2\,R_{\odot}$.  Therefore, the simple flat disk 
model predicts the circumstellar material orbiting these polluted white dwarfs is entirely contained within their Roche limits, and consistent with 
debris created via tidally disrupted asteroids.  In contrast, if the dust were not in a self-shielded configuration, 1000\,K blackbody emitters would 
lie beyond 1.5\,$R_{\odot}$ for numerous polluted stars with $T_{\rm eff}>15\,000$\,K, and debris at such distances would require a distinct 
physical production mechanism.

The flat ring model is also consistent with theoretical expectations for a disk dominated by solids, evolving at short orbital periods.  Without
gas pressure, a mature circumstellar disk orbiting a white dwarf should not have significant extent above the mid-plane.  Any primordial vertical 
height will be flattened via exchange of angular momentum between particles (i.e.\ dynamical relaxation) over many orbital periods, and these 
will be less than 3.6\,hr for material orbiting within 1\,$R_{\odot}$ around a 0.6\,$M_{\odot}$ star.  Therefore, in the absence of external forces 
a circumstellar disk orbiting white dwarf is likely to resemble the rings of Saturn.

These closely-orbiting, particulate disks around white dwarfs are likely similar to the planetary rings of the Solar System in a few important 
ways.  The first similarity is purely geometric, as the disk sizes are comparable; the outermost discrete ring of Saturn, the F-ring orbits at 
0.2\,$R_{\odot}$ above the planet (center), while the E rings extends to 0.7\,$R_{\odot}$.  The second analogy is physical, as disrupted, 
minor planetary bodies are a primary source of ring material orbiting all four giant planets \citep{esp93}, as well as the metal-enriched white 
dwarfs studied by {\em Spitzer}. Third, particulate rings evolving in isolation spread slowly due to the inefficient transfer of angular momentum 
via viscous forces among solids, and in the case of planetary rings can be further slowed by ring moons.  Therefore, similar particulate disks at 
white dwarfs can be relatively long-lived, and their evolution dominated by processes at the inner edge, where both radiation and gas production 
provide additional means to transfer angular momentum.  Fourth and finally, owing to these fundamental similarities, the observed diversity of 
rings systems in the outer Solar System is likely to be reflected in debris rings orbiting white dwarfs.

\subsection{Infrared Spectra of Disks}


\begin{figure}[ht!]
\hskip -10pt
\includegraphics[width=1.25\textwidth]{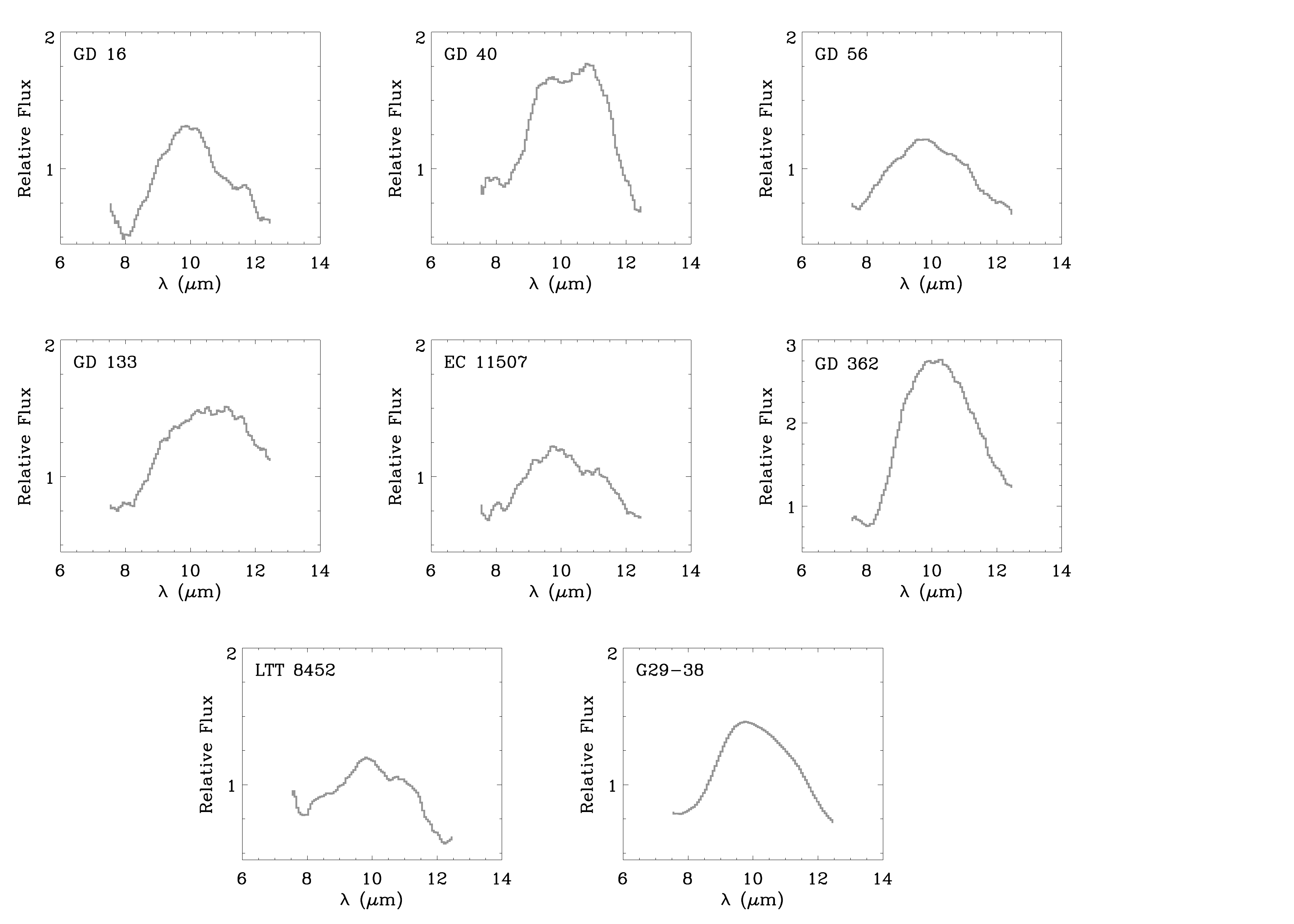}
\vskip -10 pt
\caption{Silicate emission features detected in {\em Spitzer} IRS low-resolution observations of eight dusty and metal-rich white dwarfs; 
these represent all successful $5-15\,\mu$m observations of dusty white dwarfs with IRS, and all but one star known prior to Cycle 4.  
[PG\,1015+161 was also observed but no usable signal was recovered, a likely result of the short individual exposure times imposed by 
a bright nearby source.] Each spectrum has been normalized to the average of the 5.7 and 7.9\,$\mu$m IRAC fluxes, and smoothed by 
15\,pixels (0.8\,$\mu$m); the binned data points are highly correlated and structures within the broad silicate feature are probably not real.  
The detections are modest in most cases, and the data below 8\,$\mu$m are not shown as this region is typically dominated by noise.  
Nonetheless, these data clearly show broad features, with red wings extending to 12\,$\mu$m that are typical of minerals associated 
with planet formation \citep{jur09a}.  All spectra are shown on the same scale for comparison, with the exception of GD\,362 whose 
spectacular feature requires a larger range.
\label{fig6}}
\end{figure}

Infrared spectroscopy with {\em Spitzer} IRS was successfully obtained for the bulk of dusty white dwarfs found prior to the last year of the 
cryogenic mission; the data for these eight systems are shown in Figure \ref{fig6}.  It is worth remarking on the ground-breaking sensitivity 
of these $5-15$\,$\mu$m, $R\approx100$ spectral observations obtained with a 0.85\,m telescope (in space), as the 10\,$\mu$m emission 
for all stars but G29-38 have peak fluxes in the range $0.3-1.3$\,mJy.  This can be compared with one of the most sensitive mid-infrared
observations made from the ground, where GD\,362 was photometrically detected in the $N$-band at $1.4\pm0.3$\,mJy using an 8\,m 
telescope \citep{bec05}.  Despite this ground-breaking sensitivity, all spectra but one have low to modest S/N \citep{jur09a}, and most 
disks were confidently detected only between 8 and 12\,$\mu$m where the peak in dust emission is typically 20\% to 60\% stronger 
than the $6-8\,\mu$m thermal continuum. 

Each dataset clearly exhibits strong 10\,$\mu$m silicate emission with a red wing extending to at least 12\,$\mu$m. The measured features 
are inconsistent with interstellar silicates \citep{rea05a}, but are instead typical of glassy (amorphous) silicate dust grains, specifically olivines, 
typically found in the inner Solar System and in evolved solids associated with planet formation \citep{lis08}.  In contrast, the spectra reveal 
no evidence for emission from polycyclic aromatic hydrocarbons.  These (carbon-rich) molecular compounds are found in the infrared spectra 
of the interstellar medium \citep{dra03}, some circumstellar environments \citep{slo05}, and in comets \citep{boc95}, with strong features near 
6, 8, and 11\,$\mu$m.  Hence, the circumstellar dust at white dwarfs appears relatively carbon-deficient and rocky \citep{jur09a}, and likely 
similar to the material of the inner Solar System.  These findings from infrared spectroscopy of debris in situ are consistent with, and almost 
certainly mirrored by, the carbon-poor compositions established via several metal-enriched white dwarfs \citep{jur12b,gan12,far13b}.

The only target sufficiently bright to be studied spectroscopically with {\em Spitzer} at wavelengths longer than 15\,$\mu$m is G29-38.  
\citet{rea09} obtained a second, low-resolution IRS spectrum of the prototype dusty white dwarf, this time between 5 and 35\,$\mu$m.  
Interestingly, the repeat observations over $5-15\,\mu$m revealed no discernible change in the shape of the continuum or silicate emission, 
while the longer wavelength data revealed an additional, weaker silicate feature (or combination of features) between 18 and 20\,$\mu$m.  

The opaque disk models discussed in the previous section are insufficient to account for the observed silicate emission from all dusty 
white dwarfs observed with IRS.  Specifically, the emission cannot be modeled in a disk without a vertical temperature gradient or optically 
thin region.  Completely optically thin, dust shell models have been used to reproduce the IRS data for G29-38 \citep{rea05a}; such models
are admittedly attractive as they permit the co-identification of various dust species (e.g.\ minerals, water ice) with the overall shape of the 
infrared spectra, including both continuum and emission components \citep{rea09}.  However, as discussed in detail above, optically thin 
disks are problematic for at least a few theoretical reasons, and the body of evidence on polluted white dwarfs suggests many stars have 
accreted at least 10$^{23}$\,g \citep{jur06,far10a,zuc10}, which is several orders of magnitude larger than the $\sim10^{18}$\,g of small
dust grains required for a completely optically thin disk.  These micron-size disk particles that efficiently emit at {\em Spitzer} wavelengths 
can easily be dwarfed by the mass contained in large bodies, but interior to the Roche limit the largest objects will be km-size \citep{dav99}.


\begin{figure}[ht!]
\centering
\includegraphics[width=0.5\textwidth]{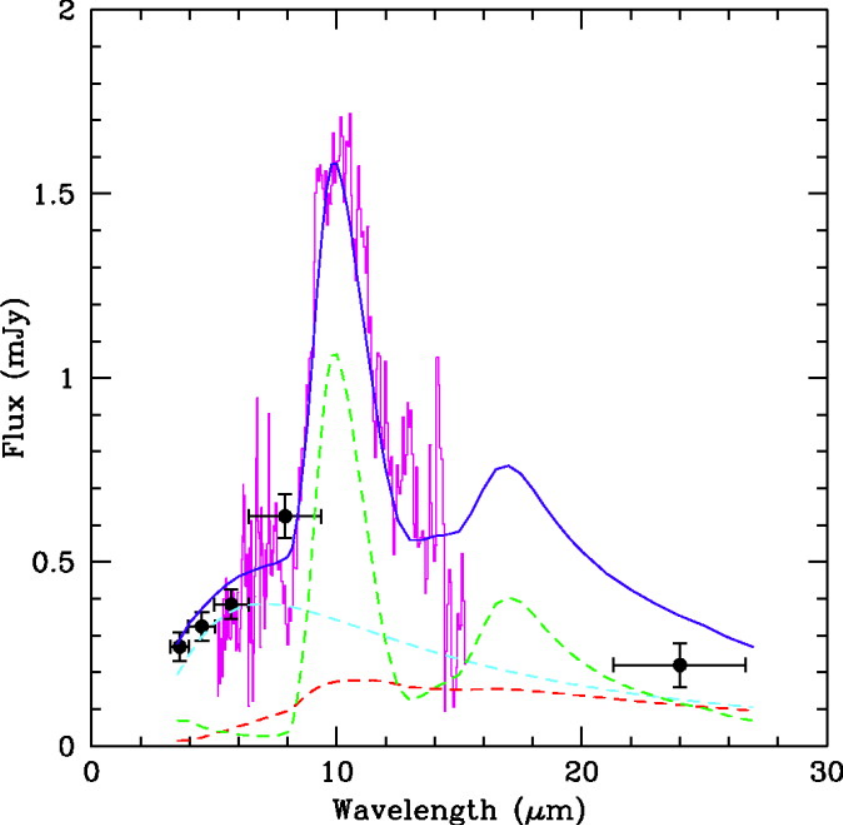}
\vskip -5 pt
\caption{A complete disk model for the thermal continuum and dust emission features at GD\,362.  Data point with error bars are photometry
from IRAC and MIPS, while the magenta line is the IRS low-resolution spectrum.  The blue line is the total model dust spectrum, which is the
sum of emission from three disk components: 1) cyan is an innermost, opaque disk; 2) red is a radially intermediate zone with a vertical 
temperature gradient; 3) green is an outermost, optically thin region \citep{jur07b}.  The only dust species used in the model are olivines.
\label{fig7}}
\end{figure}

Alternatively, the flat and optically thick disk models can be readily extended to produce an emission feature, by including either a hotter 
component on the (vertically) outermost layer, or an optically thin (radially) outer region \citep{jur07b}.  Such a model is shown in Figure 
\ref{fig7} for GD\,362, utilizing only olivine grains for both the emission components and thermal continuum.  A similar model was in fact 
used successfully for G29-38, requiring silicates only \citep{rea09}.  A further advantage of using the optically thick and geometrically thin
disks as the baseline model for infrared excess emission around white dwarfs is that these are naturally associated with a small but
significant fraction of optically thin material.  In fact, \citet{jur09b} demonstrate that for the dust properties inferred from the opaque disk 
models (e.g.\ particle size and density, opacity, disk radius), there should be $2\times10^{17}$\,g of associated optically thin material.
This compares remarkably well with the total mass of silicates invoked for G29-38 in a completely optically thin model \citep{rea05a}.

Thus, it appears the flat and optically thick models for circumstellar disks at white dwarfs provide an excellent basis for infrared emission
features beyond the thermal continuum.  The extension(s) necessary to account for the silicate emission are relatively straightforward and
well understood, but it is unclear whether these emitting regions (i.e.\ disk halos) reside in the radially outermost orbits, or at the largest 
vertical areas.  Global models of white dwarf disks indicate that optically thin material in the outermost orbital regions is rapidly removed
on PR drag timescales \citep{boc11}, implying this area must somehow be replenished with small grains to be the prime source of silicate 
emission, and thus potentially favoring an origin from the top layers of the disk.  The infrared excess at G29-38 has been present for at
least 35 years \citep{pro83}, and its strong silicate emission was likely detected in several, independent photometric measurements 26 
years ago (see Figure \ref{fig3}; \citealt{tok90,tel90}).  While not a strong constraint, together with the universally observed silicate emission,
this may support the self-sustaining production of small (micron- and sub-micron) grains that would otherwise be removed on short timescales.

\subsection{Gaseous Debris in Disks}


\begin{figure}[h!]
\centering
\includegraphics[width=0.9\textwidth]{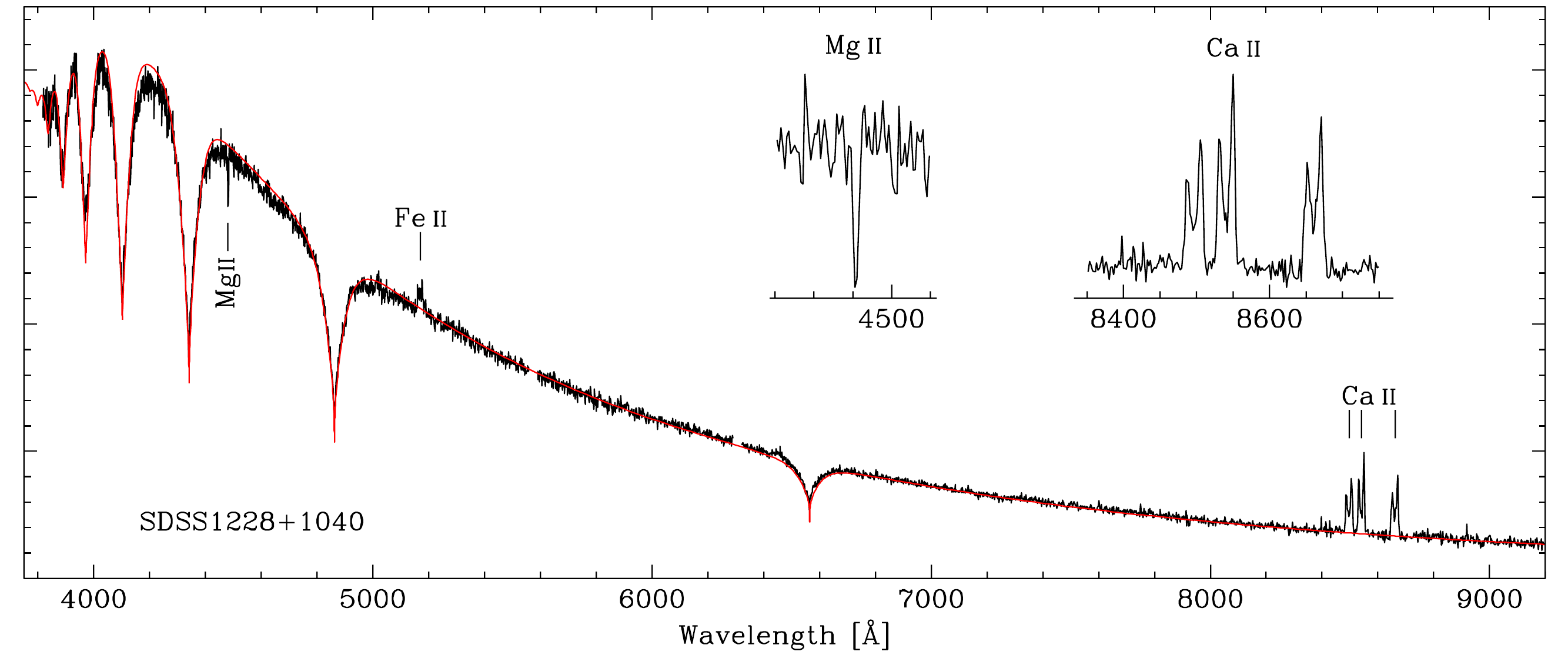}
\vskip 0 pt
\includegraphics[width=0.6\textwidth]{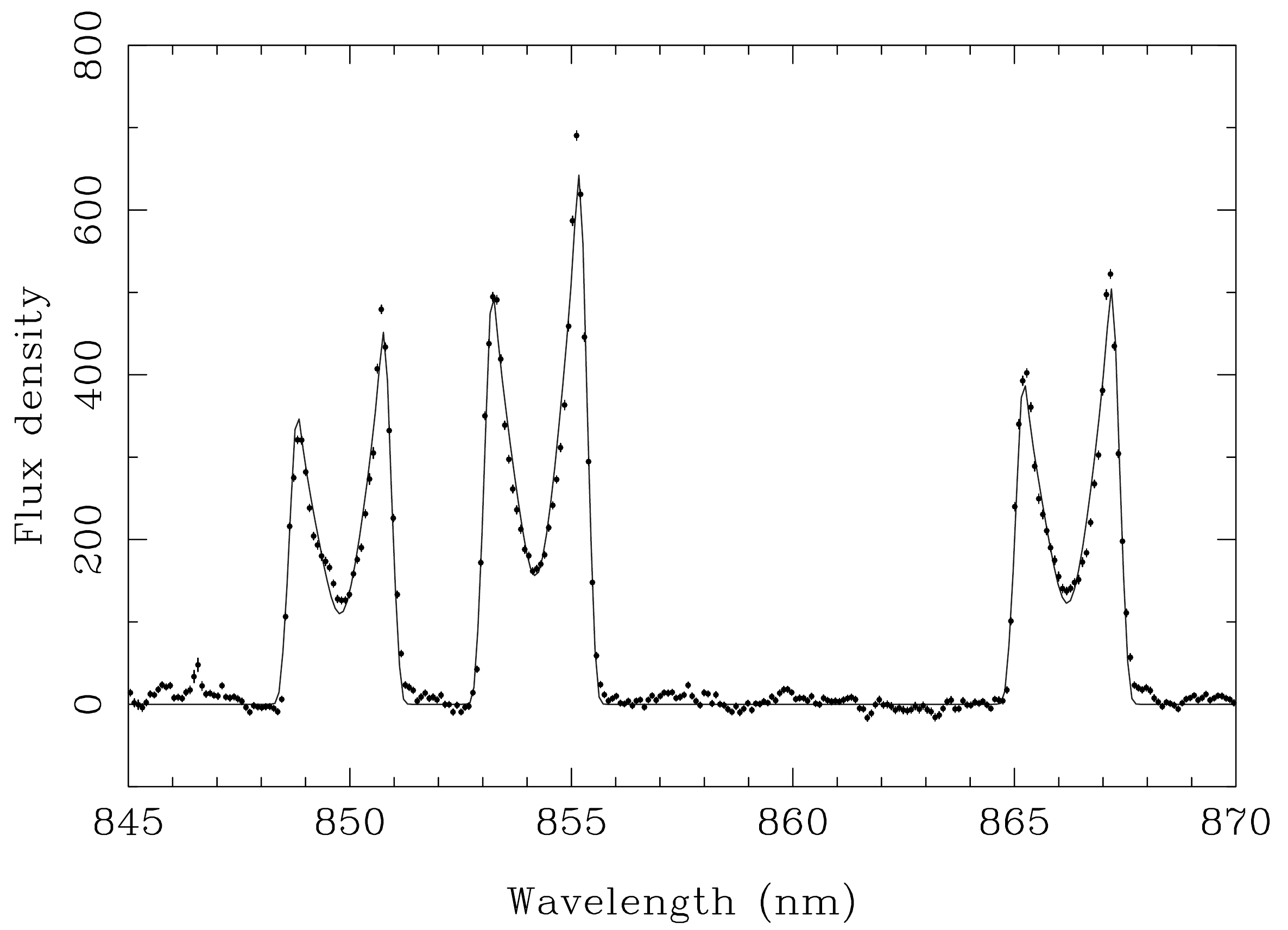}
\vskip -5 pt
\caption{Initial spectroscopic observations of SDSS\,1228.  {\em Upper panel}:  The SDSS discovery spectrum is shown in black, and 
exhibits emission lines of Ca\,{\sc ii} and Fe\,{\sc ii}, as well as photospheric absorption due to Mg\,{\sc ii}.  The red line shows a pure hydrogen 
atmosphere model, where the only expected lines are the Stark-broadened Balmer series. {\em Lower panel}:  Ca\,{\sc ii} red triplet emission 
measured at the William Herschel Telescope (WHT) with the ISIS spectrograph \citep{gan06}. The data are shown as points with (small) 
error bars, while a dynamical, disk model for the emission is shown as a solid line.
\label{fig8}}
\end{figure}

Early in the {\em Spitzer} era, and prior to the publication of any satellite observations of white dwarf disks other than G29-38, \citet{gan06} 
announced the discovery\footnote{The identification of metallic emission lines was accidental; its spectrum was flagged in a search for weak 
spectroscopic features due to very low mass (stellar or substellar) companions to apparently single white dwarfs.} of strong gaseous emission 
from a metal-dominated disk orbiting the white dwarf SDSS\,J122859.93+104032.9 (SDSS\,1228).  This remarkable object proved to be the 
prototype of a class of dusty and polluted white dwarfs with detectable gaseous debris, and this additional observational data has proven 
highly valuable in the study of disk origins and dynamics.

Figure \ref{fig8} displays a full optical spectrum of SDSS\,1228, including a high signal-to-noise observations of the Ca\,{\sc ii} triplet.  Most
remarkable are the double-peaked emission lines, whose substantial widths are the result of large radial velocity-shifts due to Keplerian rotation.
Such lines originate from orbiting, heated gas and are typical of accretion disks in cataclysmic variables and active galactic nuclei \citep{hor86}.  
However, unlike conventional accretion disks, there is a distinct lack of emission lines from hydrogen; together with the calcium emission, these 
data indicate the disk is significantly depleted in volatiles yet rich in refractory metals.  It is notable that Mg\,{\sc ii} absorption is apparent in the 
modest resolution ($R\approx2000$) SDSS spectrum of a $T_{\rm eff}\approx22\,000$\,K DA star, as photospheric metal detection typically 
requires Keck or VLT high-resolution data for white dwarfs of similar temperature and composition \citep{koe05}.  SDSS\,1228 remains among 
the most highly polluted white dwarfs known, with a calcium abundance close the solar value \citep{gan12}.

There are a few critical, observational properties that apply to the entire class of (single) white dwarfs with metallic emission lines.  The most
fundamental is that all white dwarfs with Ca\,{\sc ii} emission have detectable infrared excesses due to dusty disks \citep{bri09,bri12}; these stars 
are a subset of dusty white dwarfs and the gas component is physically and fundamentally related\footnote{There is an unfortunate tendency to 
describe stars like SDSS\,1228 as having `gas disks'; this is misleading at best, and fails to recognize the full properties of these disk systems.}.  
To date, all white dwarfs with detectable gas emission from their disks also harbor substantial atmospheric metal pollution \citep{gan08}.  This 
may be an observational bias as detection favors 1) high abundances of photospheric metals, 2) bright infrared excesses (large disk masses), 
and 3) strong emission from metallic gas.

The gaseous emission from debris disks at white dwarfs provides excellent, independent support for the conclusions of the infrared data analyses 
and the asteroid disruption model.  Interestingly, the emitting gas and dust are spatially coincident in all three cases where there are sufficient data 
to model both components in detail \citep{mel10a}.  The observed distribution of velocity shifts within the emission lines \citep{gan08} places the 
detected gas in white dwarf disks in essentially the same orbital regions as those inferred for the dust \citep{bri09}.  This is a profound {\em 
empirical} result relying only on Kepler's third law, and confirms all the material orbits well inside the Roche limit, significantly strengthening the 
model of tidally destroyed parent bodies for the disk origins.


\begin{figure}[ht!]
\centering
\includegraphics[width=0.9\textwidth]{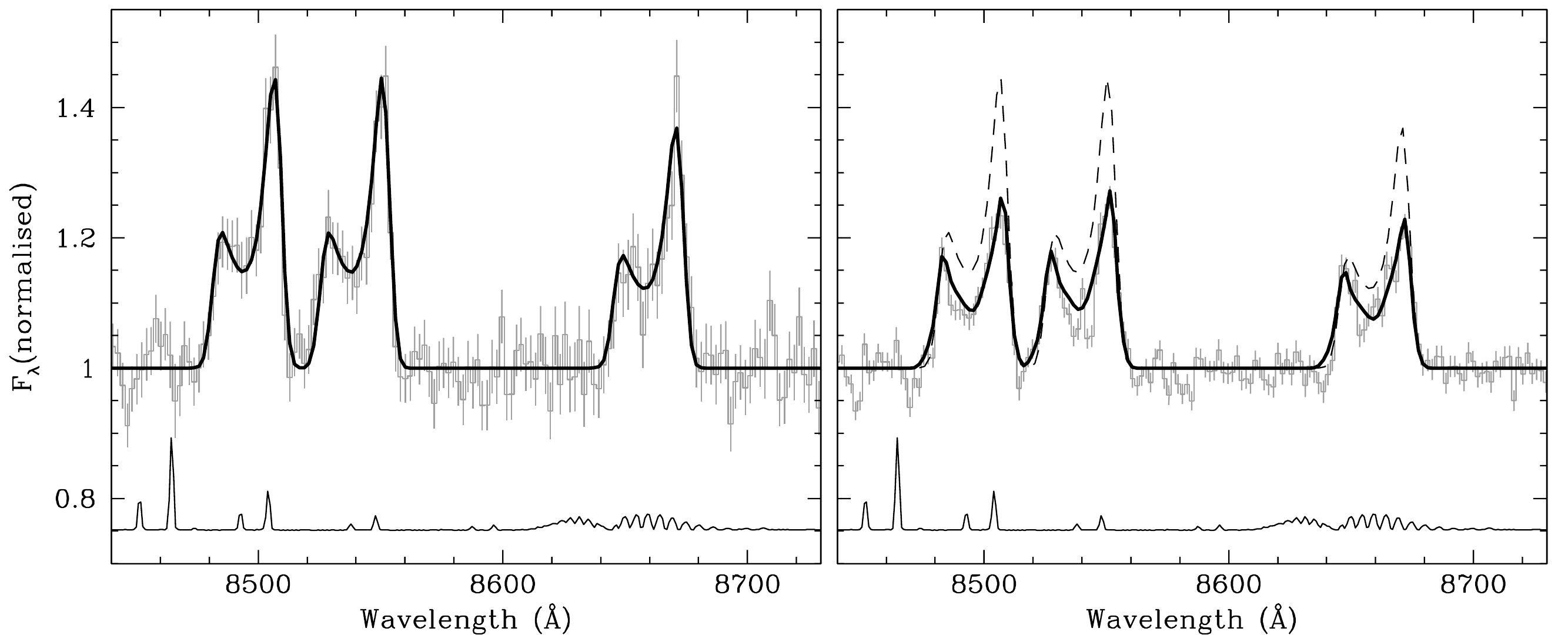}
\vskip -5 pt
\caption{The SDSS discovery spectrum of Ton\,345 taken 2004 Dec is shown on the left, and on the right is follow up data taken 2008 Jan 
with ISIS on the WHT \citep{gan08}.  The later data trace the shape of the emission lines in the SDSS spectrum, and exhibit a clear change
in both strength and shape within 3 years.  The thin solid lines at the bottom of the panels indicate the positions of strong night sky lines.
\label{fig9}}
\end{figure}

Additionally, the shape of the emission lines indicates a clear maximum velocity for the emitting gas (and thus a minimum orbital radius) 
on the order of 500\,km\,s$^{-1}$, and corresponding roughly to orbits beyond $10-20$ white dwarf radii.  These detected orbital speeds are 
distinct from the $1000-3000$\,km\,s$^{-1}$, (circular) Keplerian velocities expected for material within 10 stellar radii of the surface (and inside 
the observed disk edges); emission from such material has not been detected.  This last fact may seem surprising, since gas is expected to be 
produced via sublimation near the inner disk edges (as inferred from infrared detection of the emitting solids), spiral inward and eventually fall 
onto the stellar surface.  Hence, the behavior of the emission lines clearly indicates the detected gas is not produced from sublimation.  Further
strengthening this conclusion is the fact that gaseous emission from disks has been found for the full range of stellar effective temperatures 
where both metals and infrared detected disks are found, down to $T_{\rm eff}\approx13\,000$\,K \citep{far12a}.

Furthermore, it is noteworthy that the Ca\,{\sc ii} line profiles are often asymmetric \citep{gan08,far12a,mel12}, indicating substantial orbital 
eccentricities for the gaseous debris, and variable in shape and strength on timescales from months to years.  Figure \ref{fig9} shows the third 
disk to be discovered with metallic emission lines, Ton\,345, with two epochs of optical spectroscopy separated by 3 years that show clear 
variation in both the total flux and line asymmetries.  The possible implications of this variability will be discussed in later sections on disk 
evolution, but it is clear that observations of gas in white dwarf disks will be a highly valuable constraint on their nature.  At current count 
there are seven stars with detected emission from the Ca\,{\sc ii} triplet, photospheric metals, and infrared excesses; but only the 
circumstellar gas around SDSS\,1228 and Ton\,345 has been modeled in detail.  

For these two systems, \citet{har11,har14} constructed non-LTE, axially symmetric models of gas using a standard $\alpha$ (accretion) disk 
prescription, where the gas emission is driven by viscous heating.  The full disk components are built up from concentric rings, where each 
individual ring is fully solved for radiative transfer and structure, comprising hydrostatic and radiative equilibrium, population of atomic levels, 
metal-line blanketing, and irradiation by the central object.  These models are relatively successful at reproducing the broad morphologies 
of the Ca\,{\sc ii} triplet line profiles, including asymmetries due to eccentric rings.  If accurate, these models may constrain the chemical 
composition of the debris; the lack of observed C\,{\sc ii} emission just redward of the Ca\,{\sc ii} triplet is inconsistent with a chondritic mass 
fraction of carbon \citep{har11}, but consistent with the terrestrial value \citep{har14}.  The models also yield a strong upper limit to the H mass 
fraction of 1\%, consistent with volatile-depleted parent bodies, and finds innermost radii for the gas at modestly wider orbits than those found 
by previous studies \citep{gan08,mel10a}, supporting previous conclusions that the gas and dust coexist spatially.

Caution is warranted in the above interpretations, however, as a major weakness of the accretion disk model are strongly over-predicted 
emission lines of Ca\,{\sc ii} H \& K, and additional metal lines (e.g.\ Mg\,{\sc ii}) that are not observed.  In contrast, \citet{mel10a} modeled 
the circumstellar gas as ultraviolet-photoionized regions of metals, or Z\,{\sc ii} regions analogous to H\,{\sc ii} regions.  The numerical 
predictions of the Z\,{\sc ii} model yield total emergent fluxes from emitting gas that are in broad agreement with those observed from the 
Ca\,{\sc ii} triplet alone.  These models therefore overcome the main difficulty encountered by the viscous disk model.  It is noteworthy 
that all models for gas origins and behavior in white dwarf disks predict similar gas temperatures, and therefore scale heights of at least 
$10^7$\,cm \citep{met12}.  This clearly implies the bulk of emitting gas and dust are vertically distinct, making their radial coincidence 
remarkable if the two components are physically decoupled.  These issues are discussed further in \S5 on disk origins and evolution.

\section{Disk and Pollution Frequencies}

This section reviews the results of various searches for both atmospheric metals and circumstellar disks, and will attempt to place the
previous section into the proper historical and scientific context.  Notably, the \S3 focussed on the discoveries and detections, but the full 
statistical picture requires a discussion of the non-detections and observational sensitivities.  The connection between the metals and the
circumstellar disks has been established empirically by observations, and in this section are discussed physical mechanisms contributing
to this well-established link.

\subsection{Accretion Rates and History}

Before discussing observational strategies for detecting signatures of planetary debris orbiting white dwarfs, it is critical to introduce the
concept of inferred accretion rates.  In a series of pioneering papers, \citet{dup92,dup93a,dup93b} made the first efforts to understand the
presence and long-term evolution of atmospheric metals in cool white dwarfs, including both gravitational settling and episodic accretion.
They demonstrated that in a steady-state balance between accretion onto the white dwarf surface, and the downward diffusion of heavy 
elements from the base of the outermost layers that are chemically homogeneous with the photosphere, the mass fraction $X$ of any 
metal can be expressed as

\begin{equation}
X = \frac{\tau \dot M}{M_{\rm cvz}}
\label{eqn6}
\end{equation}

\medskip
\noindent
Here $\tau$ is the exponential sinking or diffusion timescale for a given metal species, $\dot M$ is the steady-state accretion rate of the
heavy element onto the white dwarf, and $M_{\rm cvz}$ is mass of the convection zone.  For a given metal, $\tau$ can vary significantly as 
a function of atmospheric temperature and composition (see below), but for a given star the sinking timescales for distinct heavy elements 
differ typically by no more than a factor of 2, and by no more than a factor of 3 at the extreme \citep{paq86}.

Not all polluted white dwarfs have significant convection zones, but at the time of the \citet{dup93a} study, all but one known case were 
helium-rich (DBZ or DZ) stars which possess large convective regions where atmospheric metals are thoroughly mixed.  The exception was 
the prototype DAZ, G74-7 \citep{lac83}, which is sufficiently cool to have a sizable convection zone similar to warmer DBZ counterparts.  This 
formalism has since been generalized to stars where radiative energy transport is important (or dominant), and $M_{\rm cvz}$ is taken as the 
mass of the outer stellar layers where any heavy elements are well-mixed; this is defined at the base of the convection zone or at Rosseland 
optical depth 5, whichever is deeper \citep{koe09}.

Using spectroscopy of white dwarf atmospheres and comparing with spectral models including various metal species, the abundance of any 
heavy element can be determined relative to the main atmospheric constituent (hydrogen or helium).  For any spectroscopically detected metal, 
its photospheric abundance by number is directly proportional to $X$ within the outer layers, and Equation \ref{eqn6} can be re-written to infer 
the (steady-state) rate of infall of an individual element, or summed to obtain the total mass accretion rate

\begin{equation}
\dot M_{\rm z} =  \sum\limits_{i} \dot M_{i} = \sum\limits_{i} \frac{X_{i} \ M_{\rm cvz}}{\tau_i}
\label{eqn7}
\end{equation}

\medskip
\noindent
Use of the above equations requires the assumption of steady-state accretion, meaning the rate of accretion and diffusion are equal, and
equivalent to a constant, downward mass flux for each heavy element at every optical depth \citep{gan12}

\begin{equation}
\rho X v = const
\label{eqn8}
\end{equation}

\medskip
\noindent
Here $\rho$ and $v$ are the mass density and diffusion velocity for a given metal at a specified optical layer, determined by atmospheric 
models and diffusion calculations respectively, and $X$ is measured as before via modeling of the spectroscopic data.  

These last two equations have profound importance for the empirical study of exoplanetary system assembly and chemistry.  They provide 
the means to calculate the mass accretion rate for a specific element within the infalling debris itself, and the ratio of mass accretion rates 
for any two heavy elements is an indirect determination of their intrinsic, relative abundance within the parent body (or bodies) of the debris 
\citep{kle10}.  Assessment of extrasolar minor planet compositions using measurements of metal abundances in white dwarfs, and the 
implications for terrestrial planet formation throughout the Galaxy, is outside the scope of this review but readers are referred to \cite{jur14}.  
While the subtopic of element to element ratios is peripheral to this review, the absolute abundances (and hence masses) of metals contained 
in both the disks and the polluted star are also critical.  

Total mass accretion rates must be estimated via extrapolation from detected major elements, or treated as a lower limit for undetected yet 
important species.  By mass, the most prevalent elements found in polluted white dwarfs are, to date, the same as those that dominate the 
terrestrial bodies of the inner Solar System: O, Mg, Al, Si, Ca, and Fe \citep{mcd00,vis13}; C and S are important only for some chondrites 
and comets; \citep{lod98,lod03}.  Of the few to several hundred known white dwarfs with atmospheric metals, accretion rates for these key 
elements are only known for a handful of stars studied in detail using ultraviolet spectroscopy with the {\em Hubble Space Telescope} and 
high-resolution, optical spectrographs on 8m class telescope like Keck and the VLT.  Although far ultraviolet studies have demonstrated that 
Si\,{\sc ii} (1260/1265\,\AA) is likely the most sensitive indicator (via absorption line strength) of external pollution in stars with $T_{\rm eff}\gtrsim
17\,000$\,K \citep{koe14}, a typical observational outcome is the detection of this single element.  Nonetheless, silicon alone is likely to be a good 
indicator of total accretion rate, as it is one of the four major, rock-forming elements.

For the bulk of white dwarfs in the optical, only Ca\,{\sc ii} (3934/3968\,\AA) is detected owing to the strength of this atomic transition at relatively 
cool stellar $T_{\rm eff}$ (the same reason the K line is the strongest absorption feature in the Sun).  For this reason, accretion rate calculations 
have been historically tied to calcium, thus allowing a diagnostic of ongoing or recent pollution across the (essentially) entire population of white 
dwarfs with metals.  Using the infall rate for calcium, one can infer the total mass accretion rate as follows

\smallskip
\begin{equation}
\dot M_{\rm z} \approx \frac{1}{A} \frac{X_{\rm Ca} \ M_{\rm cvz}}{\tau_{\rm Ca}}
\label{eqn9}
\end{equation}

\medskip
\noindent
where $A$ represents the mass fraction of calcium within the accreted material.  The early and fundamental work on accretion-diffusion 
in white dwarfs assumed calcium was accreted together with all the elements in solar abundance (from interstellar matter; \citealt{dup93b,
koe06}).  However, after the first several {\em Spitzer} disk detections, it became clear that the total accreted mass should be closer to 1\% 
solar by mass \citep{jur07a}, representing only the heavy elements.  Currently, using the handful of stars for which detailed measurements
of at least O, Mg, Si, Ca and Fe are available, the best extrapolation from calcium to total heavy element accretion rate is the bulk terrestrial 
mass fraction, giving $A\approx0.016$ \citep{all95,zuc10,far12b}.

Given the utility of the accretion rate diagnostic, it is important to examine when the steady-state equations might be valid, and the implications 
otherwise.  While the diffusion timescales for all heavy elements in all white dwarfs\footnote{Radiative forces can compete with and counteract 
diffusion for stars with $T_{\rm eff}\gtrsim25\,000$\,K \citep{koe14}, but this review is restricted to those white dwarfs where atmospheric pollution 
can be confidently inferred.} are universally orders or magnitude shorter than their cooling ages, they vary substantially as a function of effective 
temperature and between hydrogen- and helium-rich atmospheres \citep{paq86}.  For 25\,000\,K $>T_{\rm eff}>6000$\,K, typical sinking 
timescales are on the order of days to $10^4$\,yr for pure hydrogen atmospheres, and from 10 to $10^6$\,yr for pure helium atmospheres 
\citep{koe09}. If a white dwarf is observed to have photospheric metals, the probability that it is in a transient phase of accretion, and not a 
steady-state, is proportional to the diffusion timescale divided by the lifetime of the pollution event; $p_{\rm trans}\propto\tau/t_{\rm disk}$; the 
probability of being in a steady state is $p_{\rm ss} = 1 - p_{\rm trans}$.  Although disk lifetimes are somewhat uncertain (see \S5.3), broadly 
speaking the DAZ stars are relatively likely to be in a steady state, whereas the DBZ and DZ stars quickly approach probabilities that allow for 
transient accretion phases.  

For an isolated instance of a tidally disrupted asteroid, followed by the formation of a disk and its subsequent accretion onto the white dwarf, 
there will be three distinct phases of pollution in the stellar atmosphere.  During the first (early or build-up) phase, metals will accumulate in the 
outer layers of the star on timescales shorter than diffusion.  In this case, the heavy element ratios in the star and circumstellar disk are identical 
as any significant sinking has yet to occur.  After the passage of one to a few diffusion timescales, the system will approach a steady-state; in this 
condition the metal-to-metal ratios in the circumstellar disk and stellar atmosphere are related by the ratio of their sinking timescales \citep{koe09}.  
Finally, when accretion from the disk is complete, the photospheric metals will remain detectable for up to several diffusion timescales, depending 
on the observational sensitivity.  Any calculated mass fractions for elements within the debris itself are then dependent on the time between the 
end of the accretion epoch and the observations (generally unconstrained).

There remain some uncertainties in the state-of-the-art atmospheric models that supply the physical constants for accretion rate calculations.  The
behavior of white dwarf atmospheres at the lowest temperatures introduces some difficulty in re-producing extraordinarily broad and complex metal 
absorption features \citep{koe11}, and the highest (approaching liquid) densities found in helium atmospheres leads to uncertainties in the equation 
of state \citep{duf07}.  The recent and highly successful introduction of 3-dimensional atmospheric models has yet to tackle chemical diffusion, and
the eventual impact remains currently unknown \citep{tre15}.  The postulated existence of thermohaline instabilities in accreting white dwarf systems, 
which would potentially increase the inferred accretion rates by orders of magnitude \citep{dea13}, has been independently examined and found to
be unimportant, especially in the steady state \citep{koe15}.  From a purely empirical point of view, it is noteworthy that derived accretion rates for 
all DA white dwarfs fall into a broad but single range; stars with temperatures from 5000 to 25\,000\,K, cooling ages of 20\,Myr to 2\,Gyr, diffusion
timescales from days to 10$^5$\,yr, from purely radiative envelopes to deep convection zones \citep{koe14}.  Such consistency is unlikely to occur
if any major physics was missing from the models.

\subsection{Results of {\em Spitzer} Searches}

The first robust searches for circumstellar dust around white dwarfs were carried out with {\em Spitzer}, using IRAC.  During its cryogenic lifetime,
the instrument was able to image using two distinct cameras simultaneously; one at either 3.6 or 4.5\,$\mu$m, and the other at 5.7 or 7.9\,$\mu$m.
The ground-based, photometric studies of G29-38 and GD\,362 provided strong indications that those two disks emitted the bulk of their radiation at 
IRAC wavelengths, making it a promising tool to search for other systems in this class.  Additionally, while these first two examples (and later a few 
other prominent systems) were discovered from the ground, only one other dusty white dwarf has ever been detected from the ground beyond the 
$K$ band (GALEX\,1931; \citealt{mel11}).  Importantly, cryogenic IRAC flux measurements provided a broad and accurate characterization of the
dust temperatures and disk sizes that was not possible based on ground-based data alone.


\begin{figure}[ht!]
\centering
\includegraphics[width=0.7\textwidth]{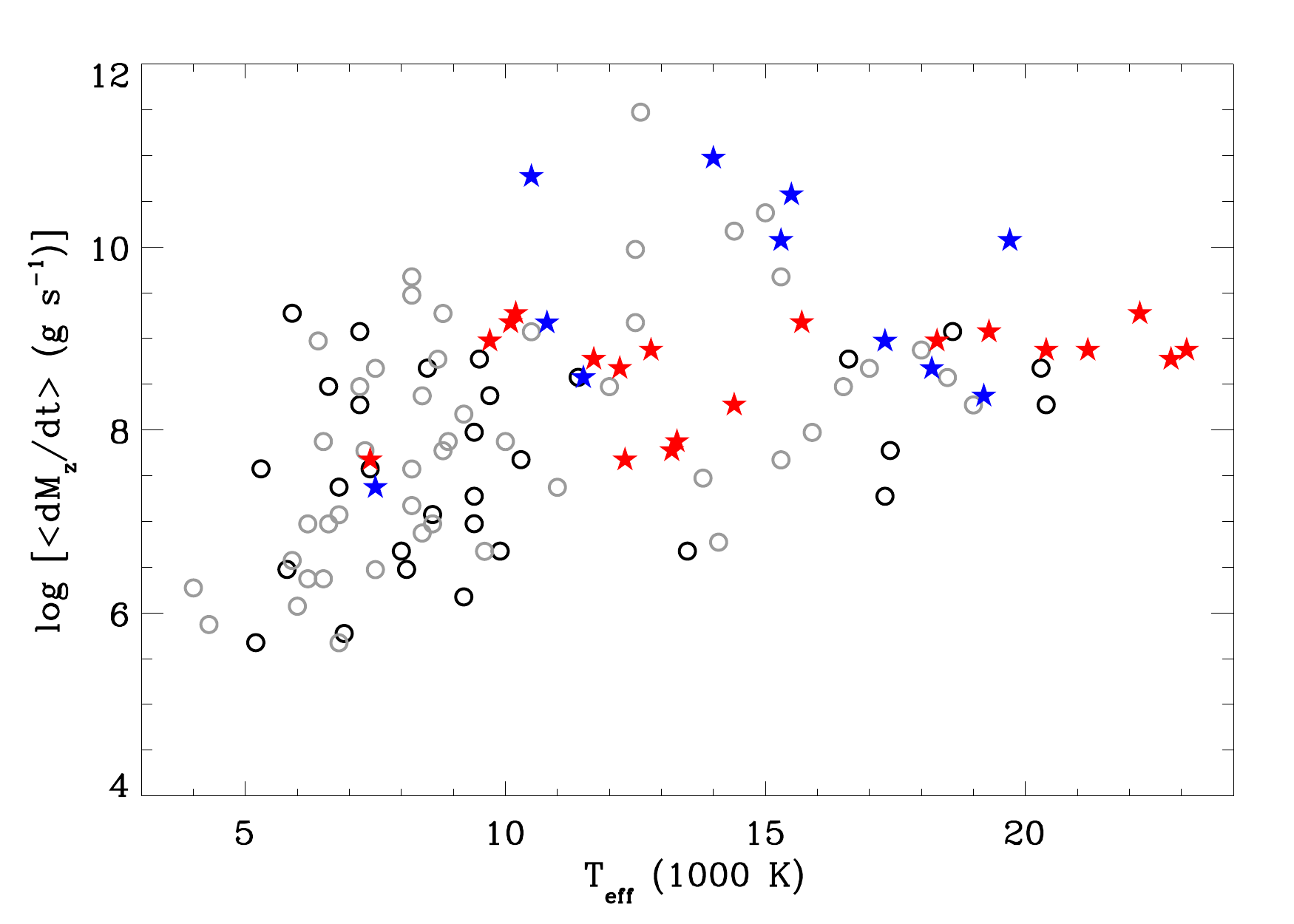}
\vskip -10 pt
\caption{Total heavy element accretion rate as a function of effective temperature for 108 white dwarfs with known (or estimated) Ca abundance 
that have been observed with {\em Spitzer} IRAC. Hydrogen-rich white dwarfs ($n=49$) with infrared excesses are plotted as red stars, and as 
black circles otherwise. Similarly, helium atmosphere white dwarfs ($n=59$) with infrared excess are plotted as blue stars, and as grey circles 
otherwise.
\label{fig10}}
\end{figure}

There were a few distinct strategies in IRAC observational searches of white dwarfs, most of which exploited the fact that both dust and 
substellar companions will produce detectable infrared excesses for a wide range of system parameters\footnote{The sensitivity of IRAC to 
spatially-unresolved, substellar companions via 4.5\,$\mu$m excess emission is competitive with state-of-the-art, extreme adaptive optics; yet 
no closely-orbiting brown dwarf or planetary companions have been discovered in well over 200 systems.}.  Early and notable programs included
searches for dust and companions orbiting a large, brightness-limited sample of white dwarfs \citep{mul07}, high-mass white dwarfs searched for 
disks due to possible stellar mergers \citep{han06,far08b}, and surveys for dust orbiting metal-rich white dwarfs \citep{deb07,jur07a,far08a}.

The basic outcome of the first wave of IRAC searches was two-fold: 1) the only stars found to have infrared excesses are those with photospheric 
metals \citep{von07}, and their fluxes are consistent with dust rather than substellar companions \citep{far08b}, and 2) many white dwarfs presumed 
to be undergoing steady-state accretion do not exhibit infrared excesses, including the DAZ prototype \citep{deb07}.  The results of these studies 
indicated that, for those systems where a steady state is likely (i.e.\ the DAZ stars), the inferred accretion rate was a good indicator of stars with 
and without detectable infrared excess.  It remains the case that white dwarfs with $\dot M_{\rm z} \gtrsim3\times10^8$\,g\,s$^{-1}$ are likely to 
have disks detectable with IRAC, whereas below this value infrared excesses are rare \citep{jur07a}.  In a basic sense, this finding appears 
physically sound if the mass (or surface density for optically thin cases) of a circumstellar disk is related to the calculated accretion rate onto
the stellar surface \citep{raf11a}.

Helium-rich stars with metals (e.g.\ types DBZ and DZ) were initially not targeted in searches for dust disks, and accretion rate calculations 
for these systems were avoided due to the uncertain, requisite assumption of a steady state.  But in 2007, the DBZ white dwarf GD\,40 was 
confirmed to have infrared excess by {\em Spitzer} based on a modest, photometric $K$-band excess in 2MASS data \citep{jur07a}, and GD\,362 
was shown to be helium-rich \citep{zuc07} even though its optical spectrum has a DAZ morphology at most spectral resolutions \citep{gia04}.  
Subsequent disk searches with {\em Spitzer} began to include all known types of polluted white dwarfs, and attempted to analyze all the stars 
on the same basis of inferred accretion rate \citep{far09}.  Within a few years, the number of disks orbiting helium-rich white dwarfs more than 
doubled \citep{far10b}, and currently there are at least ten similar systems.

The right-hand side of Equation \ref{eqn7} has a numerator that is simply the mass  of a given metal in the outer layers of the white dwarf, as 
determined by observations of spectral line strength; the steady-state accretion rate is this mass divided its rate of sinking.  This formulation 
yields an {\em instantaneous}, ongoing rate for those stars actually in accretion-diffusion equilibrium, while for stars that are unlikely (or uncertain) 
to be in a steady state, the equation provides a {\em historical}, average rate over a single sinking timescale \citep{far12b}.  In the simplistic case
of constant accretion over disk lifetime or diffusion timescale (whichever is shorter), both the instantaneous and historical rates among the DAZ and 
DBZ stars, respectively, should be comparable in the early or steady-state phases \citep{far09,wya14}.

Figure \ref{fig10} depicts infrared excess detections and non-detections, as a function of instantaneous or average accretion rate, for all the 
polluted white dwarfs observed with {\em Spitzer} during its first seven cycles \citep{ber14}.  This sample of 108 metal-enriched stars includes 
the bulk of objects known prior to or independent from those found in the Sloan Digital Sky Survey, and represents nearly all polluted white 
dwarfs specifically targeted in disk surveys \citep{von07,jur07a,deb07,far08a,far09,far10b,xu12,gir12,ber14}.  The figure makes evident a 
few remarkable properties of this sample, and of polluted white dwarfs in general, some of which will be discussed further in \S5.

\begin{figure}[ht!]
\centering
\includegraphics[width=0.7\textwidth]{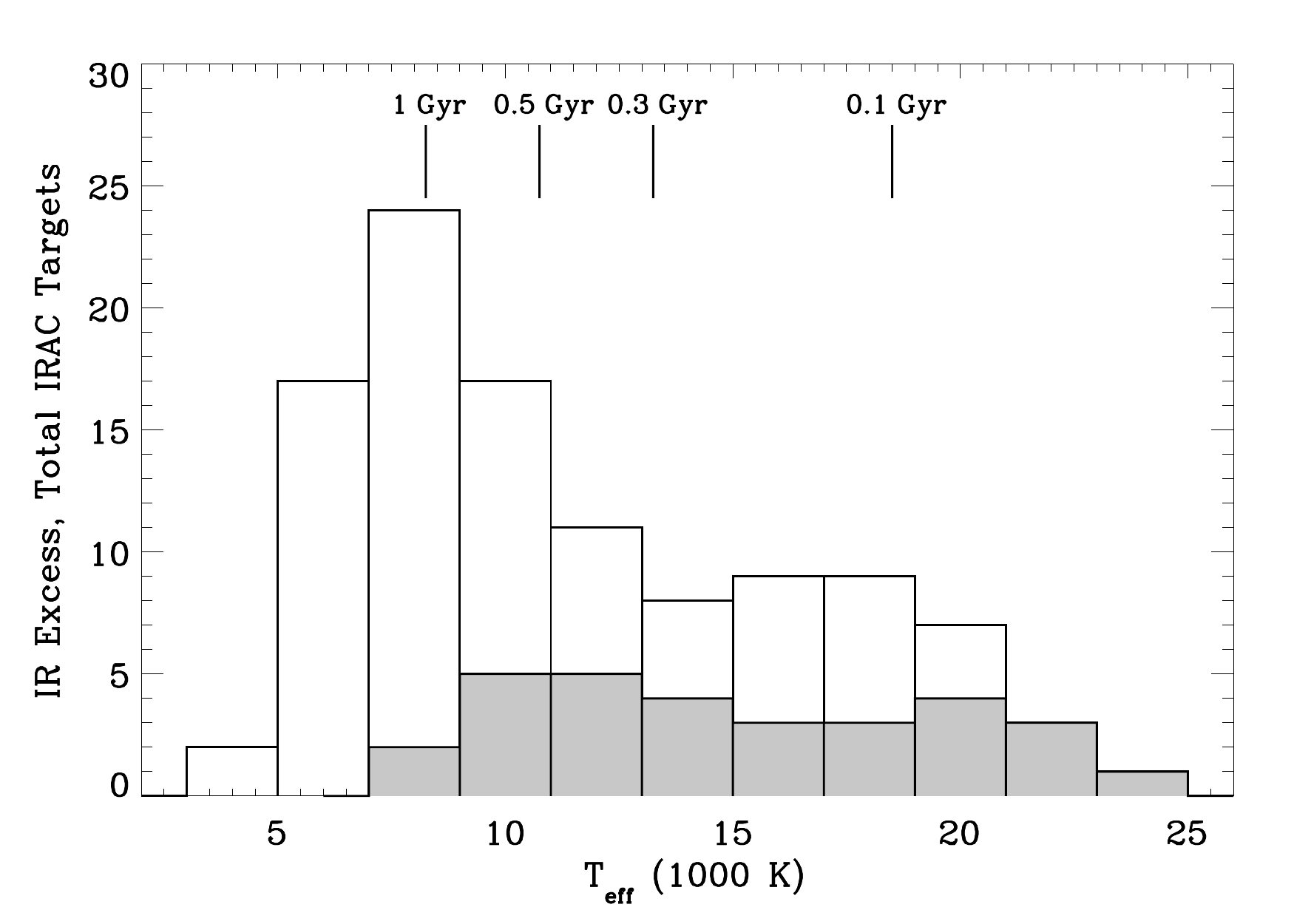}
\vskip -10 pt
\caption{Histogram of the 108 metal-rich, white dwarf {\em Spitzer} targets plotted in Figure \ref{fig10}.  The total number of objects in 
each effective temperature bin are shown by unfilled rectangles, while those with infrared excess are shown as filled, grey rectangles.  A 
few representative cooling ages are marked along the top of the figure.  While biases in the detection of atmospheric metals likely inflate the 
apparent 100\% fraction of infrared excesses around the warmest white dwarfs, there is a clear decrease in the fraction of disk detections for
older post-main sequence planetary systems.
\label{fig11}}
\end{figure}

First, the majority of infrared excess detections occur for stars around 10\,000\,K or warmer, and cooling ages less than around 0.5\,Gyr 
\citep{far09}.  Moreover, among the targets there is a decreasing fraction of detected disks with increasing cooling age, and only two possible
outliers clearly below 10\,000\,K -- where the majority of objects are located \citep{ber14}.  This result is better visualized as a histogram plotted 
in Figure \ref{fig11} for all targets as a function of effective temperature, together with the number in each bin that exhibit an infrared excess.  At
the warmest end, where cooling ages are less than around 50\,Myr, all known stars have detectable disks.  This fraction decreases to the range 
$0.3-0.5$ for post-main sequence ages of $50-500$\,Myr, and then rapidly moves toward zero for cooling ages approaching 1\,Gyr and beyond. 


Second, the sample of known polluted white dwarfs contains a strong metal-detection bias as a function of effective temperature.  That is, there
are few {\em Spitzer} targets in Figure \ref{fig10} with low to modest accretion rates and relatively warm temperatures, yet numerous examples 
at the cool end.  This property is shared by the entire population of stars whose photospheric metals are detected from the ground, and is related 
to the atmospheric opacity and the ionization balance for the Ca\,{\sc ii} H \& K doublet, both of which are a strong function of $T_{\rm eff}$ 
\citep{koe06}.  In \S2 it was mentioned that helium-rich atmospheres are relatively more transparent than their hydrogen-rich counterparts, and 
together with their longer metal residence timescales, resulted in the first dozen or so polluted white dwarfs being DZ or DBZ \citep{sio90b}.  But 
it is also the case that metals are more readily detected at cooler white dwarfs in general, and only the most dramatic cases are detectable from 
the ground above 20\,000\,K \citep{zuc03,koe05}.

Therefore, this detection bias favoring extreme cases of pollution among the warmest white dwarfs (where radiative levitation is unimportant),
combined with the empirical correlation between accretion rate and frequency of infrared excess, results in a high fraction of warmer stars with
disk detections.  This conclusion also partly accounts for the dearth of infrared excesses among polluted stars with $T_{\rm eff}<10\,000$\,K,
as the ongoing accretion rates inferred for this subset are generally low to modest. 

Third, in Figure \ref{fig10} the sub-samples of helium- and hydrogen-rich white dwarfs overlap significantly in both parameters, yet the former 
clearly extend to higher accretion rates not represented by the latter.  There are also a sizable number of DBZ stars with high inferred accretion 
rates that do not reveal infrared excesses, of which there are few among the DAZ types \citep{gir12}.  These trends indicate that the short diffusion 
timescales predicted for DA stars are corroborated by the data, where detectable disks and high, ongoing accretion rates are strongly correlated.
And based on the weaker correlation for DBZ white dwarfs, their time-averaged accretion rates are likely to reflect their actual, ongoing rates in 
only 1/3 to 1/2 of cases; for the remaining stars the current accretion rates are likely to be significantly lower.


\begin{figure}[ht!]
\centering
\includegraphics[width=1.0\textwidth]{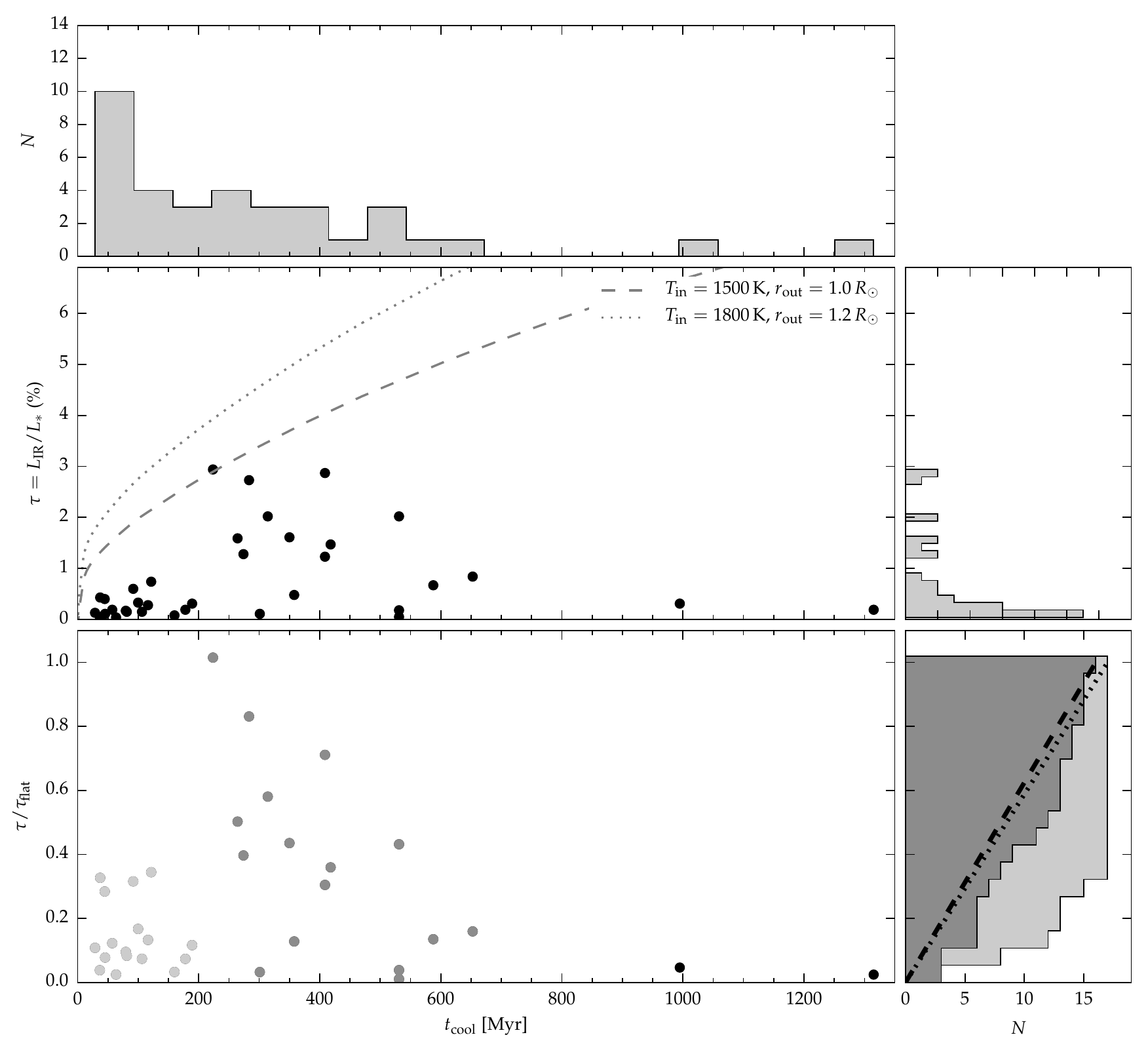}
\vskip -10 pt
\caption{Fractional dust luminosities ($\tau=L_{\rm IR}/L_*$) for all known white dwarfs with infrared-detected disks \citep{roc15}.  The central 
panel shows $\tau$ as a function of cooling age for $\log\,g=8$; the dashed and dotted lines trace the predicted values for face-on, opaque flat 
disks with radial extents given by assuming dust grains persist up to 1500 and 1800\,K, and extend out to 1.0 and 1.2\,$R_{\odot}$, respectively.  
At the top is the histogram of detections as a function of cooling age, while the center-right panel indicates these as a function of $\tau$.  The 
bottom panel plots the ratio of $\tau$ to a predicted maximum ($\tau_{\rm flat}$) for $T_{\rm in}=1500$\,K, $r_{\rm out}=1.0R_{\odot}$.  The light 
grey symbols represent $t_{\rm cool}<200$\,Myr, while the dark grey symbols correspond to 200\,Myr $<t_{\rm cool}<700$\,Myr.  At bottom-right 
are cumulative histograms of $\tau/\tau_{\rm flat}$ for the light and dark grey target subsets, and the distributions for random inclinations are 
shown as dotted and dashed lines, respectively.
\label{fig12}}
\end{figure}

Thus the statistical picture from the above {\em Spitzer} sample is somewhat distorted by target selection bias, actual accretion rate, and the 
capability of IRAC photometry to detect circumstellar dust.  Regarding sensitivity, the relevant quantity for a detection of spatially-unresolved 
infrared excess is the absolute calibration accuracy, including instrumental limitations and uncertainties in photospheric models \citep{wya08}.  
The IRAC instrument team report $1\sigma$ errors of 2\% in each of the four photometric channels \citep{rea05b}; however, photometric 
observations of isolated white dwarfs with $T_{\rm eff}>7000$\,K (and Rayleigh-Jeans emission at IRAC wavelengths), results in deviations 
from model extrapolations \citep{tre07}, and within flux ratios between channels \citep{far08a}, both of which vary by at least 5\%.  Imperfect 
extrapolation of stellar models to longer wavelengths is also a significant source of uncertainty for infrared excess searches, as many (pre-SDSS) 
white dwarfs lack reliable photometry at any wavelength \citep{mcc99,far09b}.  For these reasons most IRAC fluxes for white dwarfs are reported 
with a 5\% calibration error, which is then the limiting factor in any disk search with sufficiently high signal-to-noise and accurate short-wavelength 
fluxes.  

The most subtle excesses confidently identified at white dwarfs, using IRAC are in the range 30\%--45\% above the predicted photosphere at
both 3.6 and 4.5\,$\mu$m \citep{far09,far10b,ber14,roc15}.  This can be contrasted with a strong infrared excess like that observed at G29-38,
where the disk fluxes are a factor of $5-10$ stronger than the stellar component at the same two IRAC channels.  Also in some contrast to the 
norm, of the 35 disks detected by {\em Spitzer} to date, there are three cases where an excess is present, but only at one or more of the longest
available IRAC wavelengths: G166-58 exhibits excess at 5.7 and 7.9\,$\mu$m only \citep{far08a}, PG\,1225--079 only at 7.9\,$\mu$m, and 
HS\,2132+0941 only at 4.5\,$\mu$m (the 5.7 and 7.9\,$\mu$m IRAC detectors are not functional post-cryogen; \citealt{ber14}).  Thus, while at 
most 50\% of infrared-detected disks reveal themselves as excess in $K$-band photometry or spectra \citep{far10b}, a further 5\%--10\% are
also missed at 3.6 or 4.5\,$\mu$m, implying a small number of disks remain undetectable with warm {\em Spitzer} in cycles 6 and beyond.

Interestingly, it appears that disks with the lowest fractional infrared luminosity are the most numerous.  This should be expected if the disk
population is represented by a normal distribution in terms of intrinsic brightness, as random inclinations will not drastically alter the observed 
population.  Figure \ref{fig12} demonstrates this result and more, summarizing eight cycles of {\em Spitzer} disk observations \citep{roc15}.  
The central quantity plotted is the fractional infrared (or disk) luminosity, and it is shown for all 35 known dusty white dwarfs as a function of 
cooling age,with adjacent number histograms of detection in each parameter.  The top histogram reiterates the result shown in Figure \ref{fig11} 
and discussed earlier (in terms of cooling age rather than $T_{\rm eff}$), while the central-right panel demonstrates the bulk of detected disks 
are at the faint end of $\tau$, with the number increasing strongly towards the limit.  The two panels at the bottom of the figure show that 1) in 
contrast to observational bias, disks are preferentially fainter around younger, more luminous white dwarfs, and 2) the number of disk detections 
as a function of $\tau$ cannot be the result of inclination (at high significance for the younger population).  Together, the frequency and brightness 
of circumstellar disks around white dwarfs suggest a larger population that remains as yet undetected \citep{ber14,roc15}.  

\subsection{Independent Studies and Infrared Statistics} 

While {\em Spitzer} studies have transformed the picture of metals, accretion, and disks around white dwarfs, the sample biases are severe, 
with numerous targets already known to have strong metal pollution or ground-based infrared excess or both.  There was a single exception 
during the cryogenic lifetime of IRAC, where data at 4.5 and 7.9\,$\mu$m only were acquired for a sample of 122 white dwarfs (and 2 subdwarfs) 
based on their $K_s$-band brightness in 2MASS \citep{mul07}.  While this sample of stars is unbiased in terms of atmospheric composition, 
it favors white dwarfs with higher effective temperature and lower mass (larger radius), and the resulting sample is highly heterogeneous.  
Regardless, this is the only {\em Spitzer} study to probe such a large and diverse sample, with two disks identified (among 20 targets known 
to have atmospheric metals), and thus yielding a nominal fraction of 1.6\% with detectable disks.

Considering only metal-rich {\em Spitzer} targets, and statistically re-inserting these back into the larger, parent samples to which they belong, 
also provides an estimate of the detectable disk frequency for white dwarfs in general, resulting in a fraction of 1\%--3\% for cooling ages less 
than around 0.5\,Gyr, with an uncertain but clear decrease at older ages \citep{far09}.  If one accounts for the youngest cooling ages represented 
in the {\em Spitzer} and parent samples of stars, then it is probably safe to say this result applies to cooling ages of 20 to 500\,Myr, as neither 
metal pollution nor disk detections have yet to occur for hotter white dwarfs \citep{koe14}.

A numerically more robust approach was taken by \citet{gir11}, by cross-correlating 16\,785 DA white dwarfs with $T_{\rm eff}>8000$\,K from
the SDSS with the UKIRT Infrared Deep Sky Survey \citep{law07}.  This resulted in the identification of 1884 stars with both $H$- and $K$-band 
data whose $ugrizH$ photometry could be fitted with spectral models to search for excess emission in the $K$ band, representing an order of 
magnitude increase over all white dwarfs targeted by {\em Spitzer}.  The post-main sequence age range in this sample was based purely on the 
fact that an ultraviolet excess is necessary for white dwarf selection within the SDSS.  Twelve disk candidates were found in total, but follow 
up observations with {\em Spitzer} IRAC revealed that only 60\% of these had infrared fluxes consistent with disks, and more detailed optical 
spectroscopy demonstrated these same stars also had atmospheric metals \citep{far12a}.  The remaining candidates were photometrically 
confused with background sources in the infrared, or their data indicated low-mass companions rather than disks.  By assuming that roughly 
half of all disks are not apparent as $K$-band excess, these studies together suggests that 0.8\% of white dwarfs with cooling ages less than 
1\,Gyr have detectable disks.  Also noteworthy is that no disk candidates were found among several hundred stars with $T_{\rm eff}>25\,000$\,K 
\citep{far12a}.

By far the largest and least biased survey of white dwarfs in the infrared was performed by {\em WISE}, which imaged the entire sky at 3.4, 
4.6, 12, and 22\,$\mu$m \citep{wri10}.  Auspiciously, the first two bandpasses are relatively close matches to the two shortest wavelength IRAC 
channels, and thus direct comparisons between instruments provides verification and eases identification of new disks.  Despite the all sky
capacity of {\em WISE}, there is not yet a commensurate catalog of white dwarfs (pending {\em GAIA} results).  Thus the WIRED team opted
to cross-correlate the infrared database with each of the two largest available white dwarf catalogs: a preliminary version of the SDSS DR7 
(spectroscopic) white dwarf catalog \citep{kle13}, and the older but historically invaluable catalog of \citet{mcc99}.

In the first case, \citet{deb11a} found a detection in at least one {\em WISE} band (nearly always 3.4\,$\mu$m) for 1527 sources associated 
with a spectroscopically confirmed SDSS white dwarf.  Due to the nature of the search, just over 1000 of the detections are known or suspected 
systems with low-mass (M dwarf) companions, and where the cool component dominates the infrared emission.  Therefore, the study first had 
to determine the achieved efficiency at detecting white dwarfs, and to what flux density.  The reported $5\sigma$ sensitivity at 3.4\,$\mu$m is 
80\,$\mu$Jy but \citet{deb11a} find the peak in source counts occurs at 50\,$\mu$Jy, and that 395 white dwarfs have photospheric emission 
predicted to be above this level.  Among their 52 disk candidates, they find six sources (three previously known, and three new) above this flux 
cutoff, and thus a nominal disk fraction of 1.5\%.  However, one of these six is a system with an L-type companion (PHL\,5038; \citealt{ste09}), 
and another was found to have a neighboring (extragalactic) source contaminating the {\em WISE} photometry \citep{bar14a}.  Therefore, this 
sample yields 1\% of white dwarfs with detectable disks, over a range of cooling ages likely identical to \citet{gir11}, for $T_{\rm eff}>8000$\,K.
While this fraction could be somewhat higher, as \citet{deb11a} also cataloged another dozen white dwarfs with indeterminate-type excesses
and photospheres predicted to be above the flux cutoff, the overall survey underscores the need for careful examination and follow up of {\em 
WISE} disk candidates, especially below the $5\sigma$ sensitivity.  The relatively large ($6''$) {\em WISE} PSF at 3.4\,$\mu$m implies that 
background contamination is a serious issue, even for relatively bright sources (e.g.\ GALEX\,1931; \citealt{deb11b,mel11,roc15}).  

\citet{bar14a} performed archival and ground-based follow up of disk candidates identified by \citet{deb11a},  Of the 52 total candidates, 
they point out six are previously confirmed disk hosts, 17 are problematic due to source blending in SDSS or UKIDSS imaging, and three are 
misclassified, leaving only half the possibilities.  They obtained independent $JH$ imaging for 16 of the remaining 26 systems, and retrieved 
archival UKIDSS images for another two (with eight stars unexamined).  The independent imaging revealed 12 white dwarfs with neighboring 
sources contaminating their {\em WISE} photometry, and should be relatively unsurprising as most targets are fainter than the 50\,$\mu$Jy 
($3\sigma$) statistical flux cutoff defined by the WIRED team.  Four stars reveal no additional sources in their $JH$ images, and similarly for two 
others in UKIDSS, leading \citet{bar14a} to conclude these are all bonafide disks, adding to the six previously known for a total of 12.  However,
they do not re-calculate the number of target photospheres to which {\em WISE} was sensitive down to the faintest of their newly claimed disks;
in fact, this is not possible without extrapolation because the WIRED source counts are clearly incomplete below 50\,$\mu$Jy \citep{deb11a}.
Instead \citet{bar14a} simply discard 1062 binary suspects and claim an inflated fraction of 12 detectable disks in 465 targets, rather than 12 in 
1527; the latter is more realistic since it is likely many more white dwarf photospheres could have been detected below $3\sigma$.  Therefore
an accurate fraction of detectable disks will be closer to the latter ratio, and again near 1\%.  This presumes that all of the six new disk claims 
are real (not guaranteed), but without $K$-band imaging, background galaxies with the reddest colors can be missed.

The second {\em WISE} study \citep{hoa13} used the significantly brighter source catalog of \citet{mcc99}, which contains the bulk of all known 
white dwarfs prior to the SDSS, including the nearest populations.  Although the catalog contains 2249 entries, after some years of vetting by 
the community (misclassifications, duplications, etc.), 2202 viable targets were identified \citep{hoa07}.  As discussed previously, reliable flux 
measurements at shorter wavelengths are necessary to confidently evaluate the presence or absence of infrared excesses, and thus \citet{hoa13} 
considered only a subset of 1474 white dwarfs detected in 2MASS.  After a meticulous inspection of numerous individual cases, a total of 12 disk 
candidates were found, seven of which are new discoveries.  Here again is a detectable disk fraction that is close to 1\%, although the authors 
caution that further confirmation is needed for these candidates.  While it is possible some of these brighter disk candidates are contaminated
by additional, unresolved sources, it can also be the case that more subtle, authentic disk excesses (e.g.\ PG\,1457$-$086, HE\,0106$-$3253; 
\citealt{far10b}) are often missed.  However, the \citet{mcc99} sample is likely to contain a significant fraction of white dwarfs with $T_{\rm eff}
<8000$\,K, and based on the combined {\em Spitzer} studies, will likely result in a smaller measured fraction of detectable disks.

\begin{table}[ht!]
\small
\caption{Lower Limits for PG and SPY DA White Dwarfs with Infrared Excess}
\begin{center}
\vskip -5 pt
\begin{tabular}{ccc}
\hline\hline

$T_{\rm eff}$ (K)	&\citet{lie05}		&\citet{koe09b}\\						

\hline	

$<9500$	 		&1:16 (6.3\%)		&0:60 ($<0.02$\%)\\
$9500-17\,000$	 	&3:88 (3.4\%)		&9:246 (3.7\%)\\
$17\,000-22\,500$	&3:80 (3.8\%)		&3:138 (2.2\%)\\
$17\,000-25\,000$	&5:113 (4.4\%)		&3:179 (1.7\%)\\
Total 			&9:347 (2.6\%)		&12:626 (1.9\%)\\

\hline\hline

\end{tabular}
\end{center}

{\em Note}.  Entries are based on Table 1 in \citet{lie05}, and Tables 1 and 2 in \citet{koe09b}.  Where $T_{\rm eff}$ differs, the PG value was 
adopted for consistency.  The SPY column excludes 46 DA+dM binaries as temperature determinations are not consistently available.
\normalsize

\label{tbl1}
\end{table}

Most of the aforementioned studies cover a broad range of white dwarf temperatures and cooling ages, with the exception of the {\em 
Spitzer} studies focused on metal-polluted stars.  But based on the totality of data, disk detection in the infrared appears more favorable at 
younger post-main sequence ages \citep{xu12}, but also below temperatures where dust may be rapidly sublimated.  There have been two 
studies that attempted to capitalize on this possible peak in disk manifestation over time.

In cycle 8, {\em Spitzer} was used to study a well-characterized sample of 134 relatively young white dwarfs with cooling ages in the range $20-
200$\,Myr \citep{roc15}.  The targets were extracted from the Palomar Green (PG) \citep{lie05} and SPY \citep{koe09b} DA white dwarf samples, 
where individual stars were selected for only two qualities: 17\,000\, K $<T_{\rm eff}<25\,000$\,K and $F_\lambda(1300$\AA) $>5\times10^{-14}
$\,ergs\,cm$^{-2}$\,s$^{-1}$\,\AA$^{-1}$ (this far-ultraviolet brightness constraint ensured good signal-to-noise in a parallel, cycle 18 {\em Hubble} 
COS snapshot).  Five stars in the sample were found to display clear infrared excesses, including two previously known to have detectable dust 
in the infrared, resulting in a nominal fraction of 3.7\% \citep{roc15}.  However, it is important to note that this fraction does not include known 
binaries, but these account for no more than 10\% of the PG and SPY samples and thus do not significantly alter the fraction of young white
dwarfs with infrared excess.

\citet{bar12} propose a similar frequency based on an independent $K$-band excess search, sampling stars from the PG catalog alone.  The 
study targeted apparently single DA stars with 9500\, K $<T_{\rm eff}<22\,500$\,K, noting that all but one (then) known dusty white dwarfs lie 
in this temperature range.  However, the survey utilized disparate means of infrared excess detection, none of which were applied to the bulk 
of the sample.  Of the 117 stars in total, they obtained independent $JHK_s$ photometry for 78, while for the remaining 39 they relied on 2MASS 
data alone (with most unreliably detected at $K_s>14$\,mag; \citealt{hoa07}); a subset of 41 targets were also observed with low-resolution, $0.8
-2.5\,\mu$m spectroscopy.  From these mixed data, 11 targets with suspected $K$-band excess were observed with warm {\em Spitzer} IRAC, 
confirming one dust disk.  In addition to this newly identified infrared excess, the sample included four previously known cases, yielding an 
apparent fraction of 4.3\% with infrared detectable debris disks \citep{bar12}.  There are at least two statistical inconsistencies in this result 
\citep{roc15}.  First, there are actually 145 single DA white dwarfs in the chosen $T_{\rm eff}$ range \citep{lie05}, and it is not clear how the 
117 star subset was compiled (including four previously known disks).  Second and more fundamentally, it is well established that roughly half 
of white dwarfs with infrared excess do not reveal themselves at $K$-band \citep{von07,kil08,far09b,far10b,hoa13}; therefore the frequency 
reported by \citet{bar12} should be nearly 9\% and clearly discrepant with other results.  Underscoring this fact, the infrared excesses of 
PG\,1126+026 and PG\,2328+108 are apparent in IRAC photometry \citep{roc15}, but were both missed in $K$-band spectroscopy 
\citep{kil06,bar12}.

A statistically consistent look at the entire PG catalog yields more accurate, but still nominally high, infrared disk frequencies across several 
temperature cuts.  Table \ref{tbl1} lists and compares firm lower limits for infrared disk detections among the 347 DA stars in the PG sample 
\citep{lie05}, with 626 DA white dwarfs in the SPY survey \citep{koe09b}, suggesting that the former catalog may be a statistical outlier among 
the overall white dwarf population.  However, the differences are relatively small and within statistical uncertainties.  It may be the case that the 
PG catalog has been more thoroughly probed than the SPY sample and a few more disks await discovery in the SPY sample.

These frequency constraints, derived either partially or completely based on the PG sample, are somewhat enhanced when compared to the 
various surveys discussed above that appear to converge around 1\%.  Furthermore, it is difficult to reconcile this based on stellar temperature 
or post-main sequence system age.  For example, the much larger SDSS samples of \citet{gir11} and \citet{deb11a} are likely rather similar to the 
PG stars, since both select white dwarfs based on ultraviolet excess.  One possibility is statistical variation, and the PG catalog currently is then
relatively rich in detections by (small) chance.  Another interpretation is roughly half of detectable disks have been missed in most surveys, either 
due to lack of necessary wavelength coverage, sensitivity (and calibration accuracy), or observations themselves.  A uniform and consistent 
analysis of both the PG and SPY samples using {\em WISE} data might provide further insight.

\subsection{Gas, Dust, and Metals in the Planetary Paradigm}

The previous sections have focused primarily on infrared observations, where a sizable surface area of emitting solids is necessary for the
detection of circumstellar, planetary material.  By convention, stars with excess infrared emission are said to have a disk, while those systems
consistent with bare photosphere are often described as lacking a disk.  However, this delineation (partly limited by language) is sufficiently 
inaccurate as to obscure the actual scientific picture. The Solar System has at least two extensive belts of debris, yet its analog is far from 
detectable with state-of-the-art facilities on the ground and in space \citep{wya08}.

In the same way, gas is detectable at some fraction of dusty white dwarfs, but is presumably present in all cases where there is sufficient
circumstellar material to produce an infrared excess.  Gaseous emission from the Ca\,{\sc ii} red triplet has not yet been observed in the
absence of detectable dust \citep{bri12}, suggesting that gas production is more efficient in systems with greater masses of circumstellar
material.  While there are seven white dwarfs where both dust and gas emission are detected, and roughly three dozen with infrared excesses
in total, the potentially strong emission lines favor their discovery and gas detection in white dwarf disks is significantly less common.  Based 
on the SDSS DR7 spectroscopic sample of 4636 white dwarfs, all of which were examined carefully by eye \citep{gan07}, five were found to 
have Ca\,{\sc ii} triplet emission.  Hence the fraction of white dwarfs with detectable gas and dust is close to 0.1\%, and likely representing 
less than one tenth of the detected dust disk population

Gas in circumstellar disks can also be detected via absorption along the line of sight for favorable geometries.  For example, the well known 
edge-on disk around $\beta$ Pictoris reveals both transient \citep{hob85} and stable \citep{rob00} gas components that provide critical data on 
dynamics and composition of the parent bodies whose debris is observed.  Such detections might be relevant for those polluted white dwarfs 
without appreciable infrared excess, especially in the hypothetical case of disks that are dominated by gas-phase debris \citep{jur08}.  Although 
the disks at white dwarfs are most analogous to planetary rings, major differences will occur for any substantial gas components; the particulate 
matter can remain vertically flat while gas pressure will cause any vapor to extend far from the disk plane, to roughly 1/100 of the white dwarf 
radius \citep{met12}.  Importantly, interior to some radius where sublimation of solids is rapid, all circumstellar disks orbiting white dwarfs should 
become completely gaseous.  It is not clear if the infalling gas settles onto the stellar surface equatorially, or along lines of weak (undetectable) 
magnetic fields, but the latter case might enhance chances of detection via absorption.

\begin{figure}[ht!]
\centering
\includegraphics[width=0.5\textwidth]{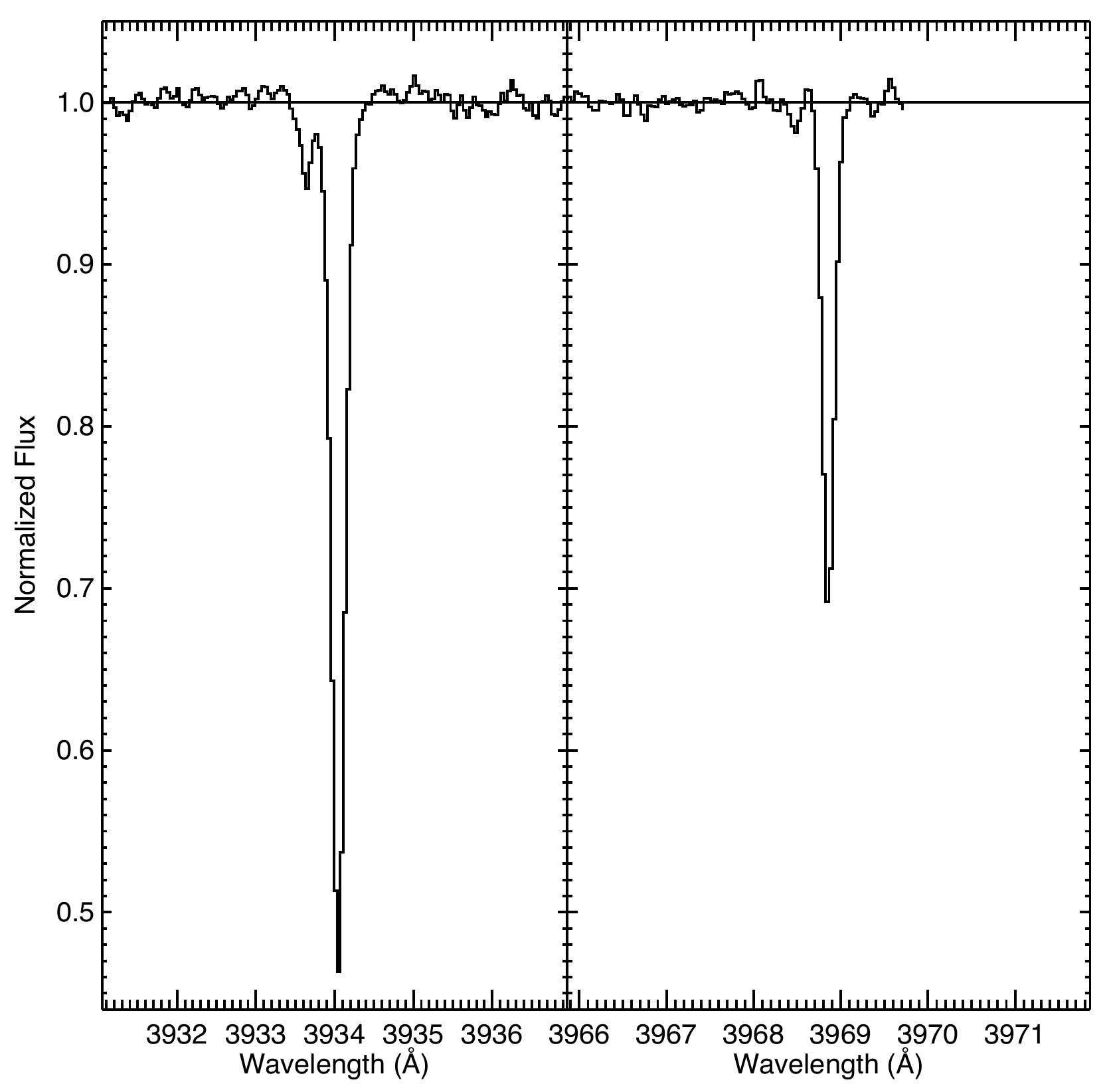}
\vskip -5 pt
\caption{An additional absorption component in the Ca\,{\sc ii} lines of EC\,11246--2923 \citep{deb12a}.  The secondary features have a velocity
shift consistent with a few to several stellar radii above the white dwarf surface, consistent with a circumstellar origin.  The relatively weak features 
were detected from the combination of numerous individual spectra taken over many nights, and totaling over 9 hours of exposure.
\label{fig13}}
\end{figure}

There are at least two, and likely three cases where absorbing circumstellar gas is implied via spectroscopic observations of polluted white dwarfs.  
Perhaps the most intriguing case is EC\,11246--2923, where the co-addition of nearly three dozen individual, high-resolution spectra taken on 11 
separate nights reveals two sets of Ca\,{\sc ii} H \& K absorption; a pair of strong photospheric features, and a set of weaker lines (Figure \ref{fig13}; 
\citealt{deb12a}). Remarkably, the secondary features are blue-shifted from the photospheric velocity by an amount commensurate with the gravitational 
well depth of the white dwarf, and are therefore consistent with circumstellar absorption.  This particular star does not exhibit an infrared excess
in cryogenic IRAC observations out to 7.9\,$\mu$m, and thus if the absorption is confirmed as circumstellar (and not interstellar), then there
are implications for disk frequency and detection.

Another two examples are for stars with strong infrared excesses, and hence the association of additional absorption components with a 
circumstellar disk is relatively confident.  PG\,0843+516 displays overly strong lines of Si\,{\sc iv} relative to a photospheric model that matches
dozens of absorption components for ten total heavy elements, including multiple lines of both Si\,{\sc ii} and Si\,{\sc iii} \citep{gan12}.  The fact 
that extra absorption is only observed for the highest ionization state, resonance lines of Si indicates a non-photospheric origin.  Moreover, the
two lines with excessive widths are blue-shifted to a velocity consistent with an orbit only a few stellar radii above the white dwarf surface.  Two 
similar but weaker analogs are observed at SDSS\,1228, where the additional absorptions are shallow with a spread of blue-shifted velocities 
relative to the photospheric lines.


Despite these promising detections of gas in both emission and absorption, it is unlikely to be detected as readily as long wavelength emission 
from dust grains.  Improved sensitivity in the mid-infrared will be realized with {\em JWST}, but for imaging studies the only improvements will 
be enhanced spatial resolution (and thus a substantial reduction in photometric confusion), and the ability to examine significantly fainter targets.
These two potential gains are unlikely to improve the scientific picture of disk properties and frequencies, as both {\em Spitzer} and {\em WISE}
are capable of targeting sufficient numbers.  However, the large mirror of {\em JWST} should enable mid-infrared spectra of polluted white dwarfs 
such that their photospheres are detected at sufficient signal-to-noise so that more tenuous and subtle excesses might be confidently identified.

By far the most sensitive indicator of remnant planetary systems at white dwarfs is atmospheric metal pollution.  In fact, all observational and
theoretical work indicates that the presence of a circumstellar disk is often (if not always) implied by atmospheric metals, regardless of infrared 
excess detection.  One particularly clear demonstration of this fact comes from the combined {\em Spitzer} IRAC and {\em Hubble} COS 
study of warm DA white dwarfs with metal diffusion timescales on the order of days \citep{koe14,roc15}.  The two programs observed 85 targets 
in common, and find at least 23 stars where accretion must be ongoing to account for the presence of atmospheric Si\,{\sc ii}, but only two of 
these systems exhibit infrared excesses.  The unavoidable conclusion is that, over the range of currently detectable metal abundances, roughly 
90\% of disks remain undetected in the infrared.

Another key result of the COS study is the metal pollution frequencies themselves, as this ultraviolet search was arguably the most sensitive to 
date.  Assuming a bulk Earth Si abundance, at least 23 of the 85 observed white dwarfs must be currently accreting debris at $\dot M_{\rm z}\sim
10^{6\pm1}$\,g\,s$^{-1}$ \citep{koe14}.  Notably, an additional 25 targets exhibit atmospheric Si that can be sustained by radiative levitation at their 
current $T_{\rm eff}$, but not within the past 50\,Myr at higher temperatures, indicating the observed metal is a recent, external contaminant.  Thus, 
the fraction of these stars with circumstellar debris is at least 27\% and possibly as high as 56\%.  These results strongly corroborate the pioneering 
studies based on optical, high-resolution spectroscopy, accounting for various differences in target properties and instrumentation \citep{zuc03,koe05,
zuc10}.  Based on the more conservative fraction, the COS survey suggests that a typical white dwarf with a cooling age of 100\,Myr accretes debris 
for around 30\,Myr, and at the observed rates would imply total accreted masses of 10$^{21\pm1}$\,g, in excellent agreement with studies of older 
polluted white dwarfs \citep{far10a}.  Such masses are equivalent to single parent bodies with diameters in the range 40--200\,km for typical asteroid 
densities.

These findings support three important scientific implications:

\begin{enumerate}  

\item{The fundamental picture of circumstellar debris and pollution at white dwarfs should recognize that, in the absence of sustaining 
mechanisms, where there are atmospheric metals there is circumstellar material.  In these cases a disk should be considered present and 
(indirectly) detected, albeit in a manner typically unavailable for main-sequence stars and debris disk in general.  Compared to metal pollution, 
infrared excess is a valuable, but significantly less efficient diagnostic of circumstellar matter orbiting white dwarfs, and identifies only the bright 
end of the disk population \citep{roc15}.}  

\item{Metals in white dwarfs are likely the most sensitive indicator of planetary debris within the inner, terrestrial zones of stars in general.  
Extrapolating from typical infrared disk detections for white dwarfs with known (and relatively high) ongoing metal accretion rates, and assuming 
disk mass scales with instantaneous accretion rate, those stars with the most modest metal pollution (i.e.\ lowest accretion rates) should be 
equivalent to disks with $L_{\rm IR}/L_*\sim10^{-6}$ \citep{far13a}.  If broadly correct, this represents two orders of magnitude greater sensitivity 
to warm dust within several AU when compared to similar detections around main-sequence stars.}

\item{The minimum disk masses implied by the accretion rates and frequencies of polluted white dwarfs are equivalent to parent bodies many 
thousands of times larger than the meter-size barrier.  Such large planetesimals are predicted as a by-product of major planet construction on 
the main sequence, and the compositions inferred (to date) support terrestrial-like analogs.  Metal pollution therefore gives the most sensitive 
lower limit for warm debris consistent with rocky planet formation \citep{koe14}, and is comparable to debris disk frequencies around young 
A-type stars that support planet formation processes beyond the snow line \citep{wya08}.}

\end{enumerate}  

\section{Disk Origin and Evolution}

This section attempts to bring together both observation and theory to reach a relatively complete picture of the detected and implied 
circumstellar disks orbiting white dwarfs.  Theoretical progress in this field has been substantial in the last several years, but the asteroid disruption 
model remains the standard.  Because disk properties have been discussed extensively above, this section follows a temporal sequence for the 
creation of disks within white dwarf planetary systems.

\subsection{Planetary System Possibilities}

While it remains to be seen exactly how often exoplanetary systems resemble the Solar System, it is an excellent starting point for extrasolar 
asteroid disruptions.  The Sun will eventually evolve into a white dwarf, after passing through the first ascent (RGB) and asymptotic (AGB) giant 
phases, and these effects have been studied by a few groups.  In brief, Mercury will be engulfed on the RGB while the orbits of Venus and Earth 
expand sufficiently due to mass loss that they escape direct engulfment on the AGB \citep{sac93}; however, tidal forces are likely to entrap and 
destroy Venus but the Earth may remain intact \citep{ras96,sch08}.  The remnant solar system should be long-term stable \citep{dun98}.  If the main 
asteroid belt is considered as one component of the inner system, then the white dwarf sun will retain 3/5 of its terrestrial components, 3/4 of the 
major planets, together with all moons and planetesimal belts.

Fittingly, the Solar System is a place of asteroid destruction.  The Kirkwood gaps of the main belt are density depletions due to orbital resonances 
with Jupiter, and objects in these regions experience chaotic eccentricity changes until they cross the orbit of Mars or the Earth \citep{moo95}.  The 
body will then gravitationally encounter a planet relatively soon, and be ejected, collide with a planet or planetesimal, or impact the Sun \citep{mor95}.  
In the latter case, if Sol instead had $R\approx0.013\,R_{\odot}$ typical of a white dwarf, an eccentric asteroid would pass inside the Roche limit and 
be torn apart by gravitational tides.  Therefore, the basic process that results in polluted and dusty white dwarfs is likely common and implies at least 
one major exoplanet.

Based on the basic idea of mean-motion (orbital) resonances, there have now been several detailed studies examining the feasibility of delivering 
sufficient planetesimal mass to the surface of white dwarfs.  \citet{deb02} was the first to recognize that although the ratio of two planetary orbits 
remains constant during stellar evolution (both expanding so that $a=a_0M_{\rm ms}/M_{\rm wd}$, typically by a factor of 2--4), their critical Hill 
separation will increase due to the reduced mass of the central star.  Hence two orbiting bodies that were marginally stable to close encounters on 
the main sequence, can become unstable during the post-main sequence.  \citet{ver13} generalized this result for two planetary bodies, demonstrating 
that instabilities can occur for orbits whose separation exceeds the orbit-crossing (Hill criteria) boundary, and for up to several Gyr into the white dwarf 
phase \citep{mus14}.

The principle of widening orbital resonances or chaotic zones was first applied to a Neptune-Kuiper belt analog carried into the post-main sequence 
\citep{bon11}.  This model predicts that a sufficient amount of planetesimal mass 1) remains available in these orbital regions at the white dwarf stage 
and 2) an adequate fraction is scattered into the inner system to account for the observed pollution, assuming a range of efficiencies for mass delivered 
to the stellar surface.  It is noteworthy that the latter process requires additional planets in the inner system to further enhance planetesimal eccentricity 
so that periastron lies inside the Roche limit \citep{bon12}, and the inner few AU region may be effectively cleared during the giant phases \citep{mus12,
nor13}.  

The strength of the extrasolar Kuiper belt model lies in the plentiful, empirical and theoretical work on main-sequence star debris disks, and which 
support these planetesimal disks retaining sufficient mass over the necessary, relatively long timescales for white dwarf systems \citep{bon10}.  
Furthermore, the collisionally evolved mass distributions expected for debris disks descending from A-type stars have been shown to be consistent 
with observed atmospheric pollution across the entire range of sinking timescales and metal abundances \citep{wya14}.  A similar argument might 
be made for Oort cloud analogs if they exist in sufficient number, with the additional benefit that their eccentricities can more easily be increased to 1 
\citep{bon12}, especially if the white dwarf is born with a velocity kick due to anisotropic mass loss \citep{sto15}.  However, it has been demonstrated 
that the Oort-like cloud mass that can reach the Roche limit of the white dwarf is insignificant when compared to the observed metal abundances, 
and hence cannot be a source for polluted white dwarfs \citep{ver14b}. 

Outer planetesimal belts similar to the Oort or Kuiper belts should be rich in volatiles, which are found to be strongly depleted in the debris orbiting 
and polluting white dwarfs \citep{jur09a}.  Were similar objects often disrupted near a white dwarf, relatively light elements such as carbon and nitrogen 
that are common in volatile compounds such as ices cannot escape the pristine photosphere \citep{gan12} by stellar luminosity, rotation, or magnetic 
field; there is no process favoring heavier or more refractory materials, yet these elements are nonetheless absent or markedly deficient relative to 
solar values \citep{jur14}.  Moreover, for bodies of radius 50\,km or greater, the bulk volatile content of Kuiper belt analogs should be retained during 
the luminous post-main sequence phases \citep{jur10}.

These compelling findings have led researchers to favor extrasolar, terrestrial-zone parent bodies as the origin of external pollution.  Exploring a 
Jupiter-main belt model extrapolated to the white dwarf phase, \citet{deb12b} find the 2:1 mean-motion resonance is sufficiently widened to perturb
and tidally disrupt a significant number of large asteroids over cooling ages of 10--300\,Myr.  Scaling from the Solar System, they find a typical system 
requires $10^2-10^4$ times the current mass of the asteroid belt so that a sufficient material is eventually accreted by the white dwarf.  Given that the
progenitors of white dwarfs are typically A-type stars, 2--3 times more massive and a few Gyr younger than the Sun, this is possible but still potentially
extreme.  Another key study simulated the influence of a 4\,AU planet on a wide belt of planetesimals spaced uniformly over 10\,AU, with a particular
focus on the effect of planet eccentricity and mass on the perturbation (and tidal destruction) efficiency \citep{fre14}.  These simulations demonstrate
that eccentric planets around half the mass of Neptune are the most effective producers of disrupted asteroids, relative to more massive giant planets, 
and their effectiveness increases monotonically with eccentricity.  

Interestingly, \citet{fre14} also find that the mass of the planetesimal disk must be a few thousand times larger than the asteroid belt, corroborating 
the findings of \citet{deb12b} and highlighting a current dilemma.  While extrasolar asteroid belts of the necessary size and mass are consistent with
observational limits, and hence not ruled out empirically \citep{wya08}, owing to collisional evolution, massive disks within 10\,AU are not expected 
to survive beyond a few Myr, and should be depleted by orders of magnitude by 100\,Myr, well before the post-main sequence \citep{wya07,bon10}.
There is currently no proposed solution to this puzzle: chemistry favors terrestrial-like parent bodies formed interior to a snow line, and planetesimal
belt evolution favors the icy, outer reaches.

It may be the case that asteroid-like belts are relatively common, as are extrasolar Kuiper belts.  From a formation point of view, a gas giant formed 
near a snow line can easily prevent the coagulation of a planet just inside their orbit, creating an asteroid belt in the terrestrial-zone \citep{mar13}.
The exoplanetary systems orbiting the A-type stars HR\,8799, Vega, and Fomalhaut all appear to share this basic architecture \citep{su09,su13} 
with asteroid belt analogs located near (and interior to) their respective snow lines, and with (confirmed or inferred) giant planets just exterior.  Early 
simulations demonstrate that the exoplanet HR\,8799e will often perturb its exo-asteroids towards the inner system \citep{con15}, suggesting that a 
progenitor architecture of polluted white dwarf systems, one which is similar to the Solar System, is not unusual.

\subsection{Formation and Structure of Debris within the Roche Limit}

This section focuses on the parent bodies and their debris as it evolves under the influence of tidal disruption and eventual accretion onto the white
dwarf surface.  A discussion on white dwarf disks may default to the debris detected within the innermost region, but planetesimal belts that supply the
destroyed parent bodies are also debris disks, albeit not yet detected for polluted white dwarfs \citep{jur09b,hoa13,far14}.  Several, more distant and
colder disk candidates have been identified around hot and luminous white dwarfs by 24\,$\mu$m excess emission, the most famous example being
the central star of the Helix nebula \citep{su07}.  However, their interpretation as genuine debris disks is complicated by binarity and dusty 
outflows associated with their immediate progenitors \citep{chu11,cla14}.   

Tidal destruction for a solid body orbiting a star has not been modeled as fully as have stellar disruptions around compact objects, but there should be 
fundamental similarities.  The catastrophic event necessitates passage through the Roche radius, which in turn implies $e\gtrsim0.998$ for semimajor 
axes beyond a few AU, and hence the orbit will be virtually parabolic.  From energetic considerations about half the mass of the fragments will become 
unbound \citep{lac82,ree88}, an important outcome that has been overlooked when inferring parent body masses based on the observed photospheric 
metals.  Bound material remains on highly eccentric orbits, at least initially \citep{eva89}; rubble pile simulations with only gravitational forces show that 
bound pieces will continue along their original orbits via increasingly filled rings, and that some particle clusters require multiple passes to be disrupted 
\citep{deb12b,ver14a}.  

Assuming only stellar radiation and gravity, it has been shown that rubble piles disrupted into collision-less rings will gradually contract under PR drag,
on timescales a few to several orders of magnitude shorter than the white dwarf cooling age \citep{ver15}.  In this picture, rings composed of identically
sized particles contract as separate units, with a spread in contraction timescales over as many decades as there are particle sizes.  This can be orders 
of magnitude longer than a few hundred orbits of the parent asteroid, and can readily exceed a few Myr for cm-size particles.  Because the known disks 
have comparable lifetimes based on both theoretical and observational studies \citep{kle10,boc11,gir12}, infrared observations are sensitive to any debris 
orbiting within a few AU but outside the Roche limit.  The non-detection of contracting rings of debris \citep{jur07a,far09} demonstrates that disks form 
significantly sooner than from PR drag alone, which specifically cannot pull the largest fragments (and hence masses) into close orbit.

Although speculative, it is likely that two additional processes are the key drivers in the formation of closely orbiting disks from a highly eccentric stream 
of debris: collisions and sublimation.  Micron-sized dust particles known to be present from infrared observations will have their orbits altered significantly 
and rapidly by radiation drag; e.g.\ 10\,$\mu$m grains with $a=5$\,AU will have their semimajor axis reduced by up to a few percent within a single orbit 
after disruption \citep{bur79}.  Therefore at periastron on subsequent passes, there will be a spread of perturbed orbital parameters corresponding to the
full range of fragment sizes immune to tidal gravity.  Keplerian velocities within the Roche radius are a few to several hundred km\,s$^{-1}$, ensuring that 
collisions will occur at high speed, and likely generate both gas and dust as smaller fragments are produced. 

Gas facilitates the viscous dissipation of energy, and is likely necessary to both compress and circularize the debris into the observed compact 
configurations.  Sublimation is critical for the eventual accretion of disk material onto the white dwarf surface, but may also be an important means 
of generating gas and viscosity within the first few orbital passes following tidal destruction.  Because the periastra of disruption events are most likely 
to occur near the actual Roche radius, a typical fragment will spend nearly one hour within this zone \citep{ver15}.  During these early stages of disk
formation, and prior to significant shielding, myriad fragments and debris will be fully exposed to stellar radiation, where they can attain surface 
temperatures above 1200\,K and sublimate small grains such as olivines \citep{deb11c}.

An interesting possibility is that periastron becomes the focal point for a collisional cascade within the first several passages, and then a dominant 
and stationary site for dust and gas production over at least a few hundred orbits \citep{jac14}.  This is precisely where most collisions are expected 
for eccentric orbits \citep{wya10}, and where any additional, star-grazing bodies would greatly enhance the collisional cross section.  In the Solar
System, similar collisional cascades following disruptions are supported for the rings of Saturn, Uranus, and Neptune, as evidenced  by the size 
distribution of the largest fragments \citep{col00}, and moonlets within the Roche limit that are too large to have formed by accretion \citep{sre07}.
If this simplified picture is basically correct, disks at white dwarfs could form within several tens of orbits of the disrupted parent body, and thus 
possibly in as little as several hundred to a few thousand years.  Assuming a Myr total disk lifetime, this relatively short formation timescale is 
consistent with three dozen disk detections in the infrared, where no dust was detected external to the Roche radius.

Major and detailed theoretical efforts have been made to understand the evolution of disks orbiting within the tidal radius of white dwarfs \citep{deb11c,
raf11a,raf11b,boc11,met12}.  To begin, it is worthwhile to summarize the possible and likely overall disk configuration and constituents.  The outer disk 
should be mainly comprised of solid debris, and while particle sizes have little empirical constraint, it is likely that micron- to centimeter-sized grains are 
the norm.  Interior to some radius where sublimation is efficient, the disk should become entirely gaseous before its eventual accretion onto the white 
dwarf.  Owing to the relatively rapid exchange of angular momentum between debris on orbital periods less than a few hours, and in the absence 
of major gravitating bodies (e.g.\ nearby planets, ring satellites), all solid state matter should lie in a geometrically flat plane, with a vertical height 
comparable to the largest disk particles.  In contrast, the inner disk will be vertically extended due to gas pressure, and hence also rotate slightly 
slower than the Keplerian speed.

The particulate disk will have typical orbital velocities of a few to several hundred km\,s$^{-1}$ between the Roche limit and some radius at which 
grains are effectively sublimated.  Despite these high velocities, and the fact that collisional timescales are comparable to the short orbital periods 
in optically thick disks, it has been shown that collisions are unlikely to be important in two ways.  First, the collisional transport of momentum (i.e.\ 
viscosity) resulting from differential rotation and random motions is orders of magnitude insufficient to account for the inferred rates of disk accretion 
\citep{far08a}, for any reasonable particle sizes and disk masses \citep{met12}.  Second, energetic collisions will be effectively damped within particulate 
disks of high surface density, and thus cannot be a significant source of gas \citep{met12}.


\begin{figure}[ht!]
\centering
\includegraphics[width=0.5\textwidth]{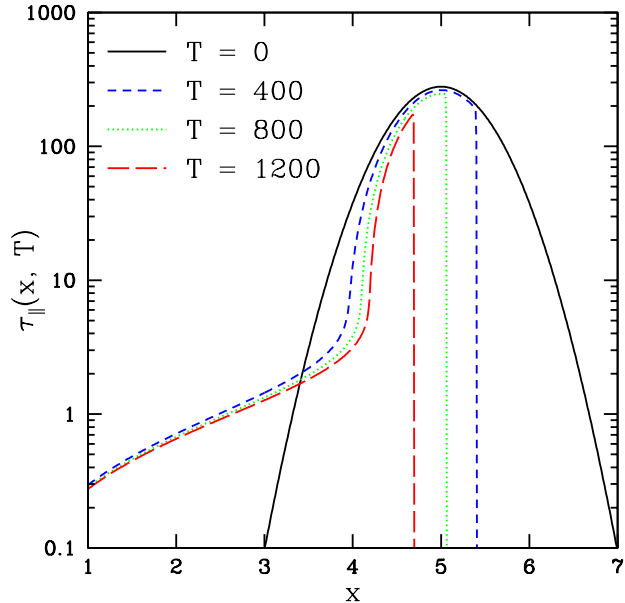}
\vskip -10 pt
\caption{The evolution of an optically thick disk with an initially Gaussian surface density profile under the influence of radiation forces \citep{boc11}.
The $y$-axis plots the optical depth as a function of orbital radius, represented on the $x$-axis as integer multiples of the inner dust disk edge.  The plot 
shows the disk profile at four time steps measured by the characteristic PR drag timescale T for cm-size particles, with two remarkable features.  First, 
an abrupt outer edge develops immediately, and marches gradually inward, re-shaping the disk from the outside even while erosion proceeds from the 
inside.  Second, there is a persistent, optically thin tail at the inner disk edge, where the PR effect operates and drives inward mass transport.  This 
behavior is generic for all initial disk density profiles.
\label{fig14}}
\end{figure}

Disk evolution and structure are strongly influenced by stellar irradiation.  The overall shape of the dusty component is determined almost exclusively 
by the radial extent, from near the periastron where the parent body is destroyed inward to some radius ($r_s$) where sublimation dominates.  Two 
extreme cases that can result in zero radial width, and likely a short-lived and completely gaseous disk, are asteroids disrupted either within 1) $r<r_s
\ll1R_{\odot}$ or 2) $r<1R_{\odot}<r_s$ \citep{von07}.  Relatively narrow rings should result when these conditions are minimally avoided (see Figure
\ref{fig12}; \citealt{far11a,roc15}).  Furthermore, all dust disks should have sharp outer edges, as any optically thin material is quickly dragged into 
regions of high surface density (Figure \ref{fig14}; \citealt{boc11}).  Inward transport of mass and momentum is driven by the PR effect at the inner 
edge of the particulate disk \citep{raf11a,boc11}, but may also be enhanced by the presence of gas \citep{raf11b,met12}.  Although only the innermost 
solids are exposed to the full starlight, the relatively rapid inward drift and sublimation of unshielded particles results in a substantially lower optical depth 
at this boundary, and thus a commensurate increase in the mass of particles drawn inward per unit time \citep{raf11a}.  

Assuming disks are dominated by dust and rubble, the inward flux of these solids driven by radiation pressure is the bottleneck for mass transport. In 
contrast, the entirely gaseous portion of the inner disk should be readily ionized and subject to the magneto-rotational instability (MRI), yielding viscous 
dissipation timescales that are orders of magnitude shorter than those resulting from PR drag \citep{jur08,raf11b,met12,far12b}.  Conservation of mass 
and this dichotomy of inward transport speeds between solid and gaseous debris, implies the inner ring of gas will have a surface density orders of 
magnitude lower than the dust disk.  This should produce infalling gas that is optically thin, and consistent with observations \citep{mel11,deb12a,gan12}.  
The final inspiral for disk material may be mediated by weak magnetic fields as low as a fraction of a kG \citep{met12}, (negligibly) spinning-up the star 
and directing the debris stream(s) away from the disk plane onto poles.  Otherwise the accretion will proceed equatorially, and may induce shocks as the 
gas orbiting near $0.01c$ comes into contact with the white dwarf surface rotating at a few km\,s$^{-1}$.  Either process may result in short-lived 
inhomogeneities in the surface distribution of metals \citep{mon08,tho10}.

\subsection{Disk Accretion and Lifetime}

The disk evolution models introduced in the previous section provide sets of predictions that can be compared to observational data, primarily via 
metal accretion rates and infrared excesses.  As a baseline for what follows, consider the models of PR drag on a flat and optically thick disk of solids 
as the baseline evolutionary process.  Figure \ref{fig14} demonstrates that this mechanism erodes the inner edge of the dust disk, but actually drives 
the entire disk structural evolution from the outside inward, while simultaneously retaining a clean outer boundary.  Therefore, rings of dust should 
narrow as they accrete onto the white dwarf, and the rate of tapering should increase with stellar effective temperature and luminosity.  Based on 
the flat disk model, infrared observations appear to support these predictions of narrower rings at younger and warmer white dwarfs, although particle 
sizes may also play a role (Figure \ref{fig12}; \citealt{roc15}). 

Notably, the speed of dissipation onto the central star by PR drag is restricted by geometry in the case of opaque rings, and for a given star depends 
only on the temperature and location of the innermost solids; in fact, the accretion rate cannot deviate significantly from a fixed value below $\sim10^9
$\,g\,s$^{-1}$; \citep{raf11a}.  In this case, the total disk mass determines its lifetime, supplying continuous accretion over at least several Myr for large 
asteroid masses.  For low-mass disks that are optically thin, the PR effect operates everywhere simultaneously and for a given star the accretion rate 
depends on the size, density, and orbital radius of particles.  Accretion rates in such disks are always lower than for their opaque counterparts 
\citep{boc11}, while disk lifetime is identical to the PR drag timescale, with likely upper bounds approaching Myr for cm-size debris at large orbital radii.

For a range of reasonable assumptions, this baseline mechanism can account for all but the highest inferred metal accretion rates.  More specifically 
and importantly, the predictions of PR drag on white dwarf disks can account for the full range of instantaneous accretion rates \citep{raf11a,xu12}, 
including all the DAZ stars with diffusion timescales less than 1000\,yr.  That is, the PR drag model successfully describes the subset of white dwarfs 
where ongoing, steady-state accretion is likely.  Moreover, the disk lifetimes predicted by these models are a good match to those inferred from 
observations based on 1) the fraction of polluted white dwarfs over a given range of cooling ages \citep{zuc10,koe14}, 2) the mass of metals in the 
outer layers of convective stars divided by typical accretion rates \citep{gir12}, or 3) the steady-state accretion phase necessary for self-consistent 
oxygen chemistry \citep{kle11}.  These combined theoretical and empirical data support a picture where the most prominent disks have 
$m\gg10^{20}$\,g and hence parent bodies with $d\gg40$\,km \citep{wya14}.


\begin{figure}[ht!]
\centering
\includegraphics[width=0.7\textwidth]{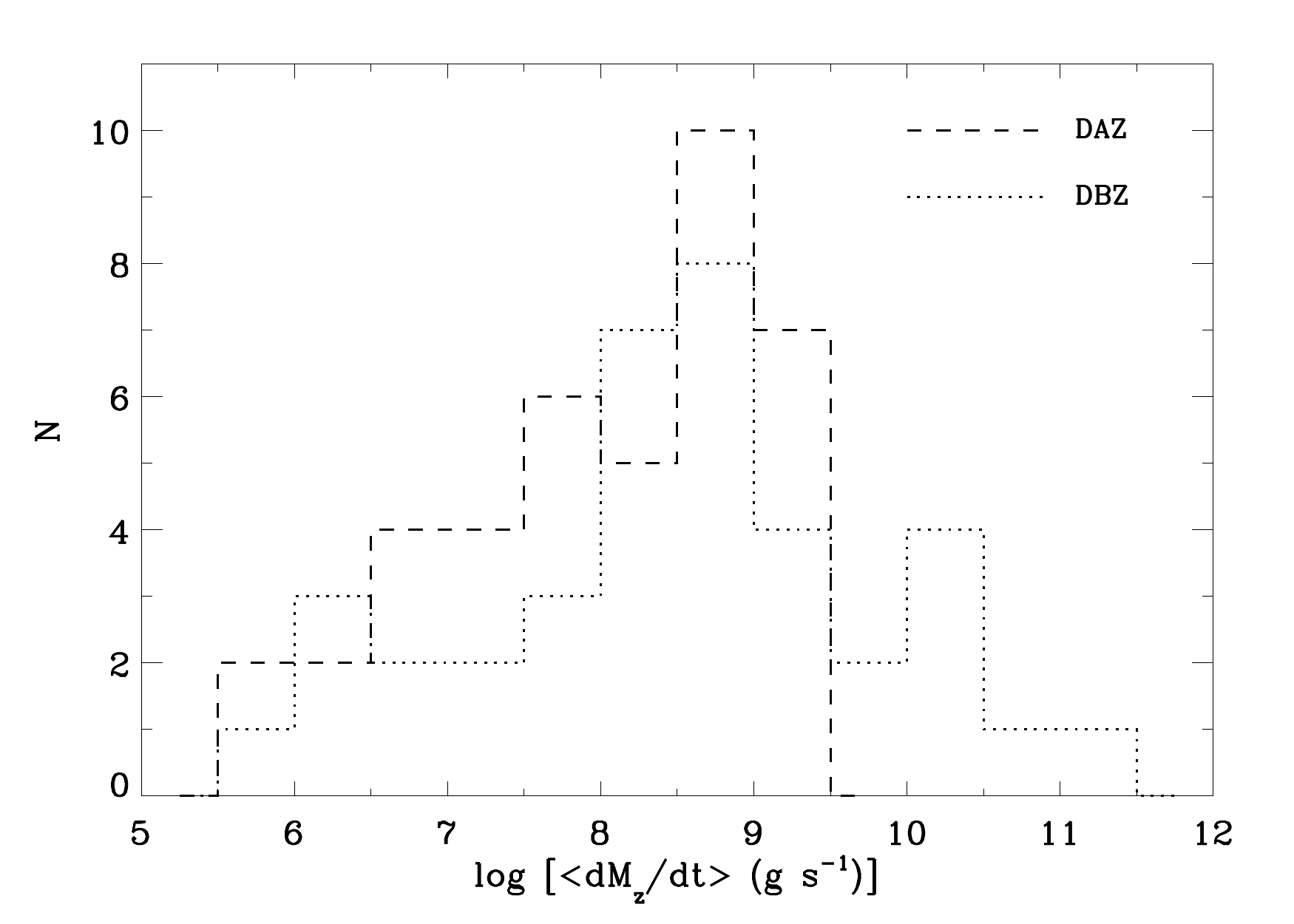}
\vskip -10 pt
\caption{Histogram of inferred accretion rates for 78 polluted white dwarfs observed with {\em Spitzer} \citep{far12b}, separating the hydrogen (DAZ) 
and helium (DBZ-type) atmosphere stars.  The population of rates exceeding $\dot M_{\rm z}>10^{9.3}$\,g\,s$^{-1}$ occurs only among the DBZ-type 
stars where only time-averaged rates can be calculated, and thus must be related to their significantly longer metal sinking timescales.
\label{fig15}}
\end{figure}

Remarkably, the highest inferred accretion rates for polluted white dwarfs -- those that are inconsistent with mass transport via PR drag -- occur only 
for those stars with helium-dominated atmospheres \citep{gir12}.  This fact is epitomized in Figure \ref{fig15}, where a histogram of accretion rates for
78 {\em Spitzer} targets, calculated assuming a steady state, reveals an upper threshold of $2\times10^9$\,g\,s$^{-1}$ for DAZ stars where ongoing 
accretion is virtually certain.  In contrast the DBZ and similar white dwarfs, where metals can persist for $10^4-10^6$\,yr, exhibit a clear subset of rates 
beyond the apparent DAZ limit by one to two orders of magnitude.  The inescapable conclusion is that some fraction (or possibly all) polluted white 
dwarfs experience short periods of high intensity accretion.  Any high-rate accretion episodes must be sufficiently brief that they have eschewed 
detection among hundreds of surveyed DA stars, and also sufficiently extreme that their present-day, time-averaged rates still bear the scars 
\citep{far12b}.

There have been a few ideas suggested to produce accretion rates exceeding those resulting from PR drag, and all involve gaseous debris.  By far 
the most detailed and developed is the model of runaway accretion \citep{raf11b}.  In this scenario, gas produced by sublimation at the inner edge of
the particulate disk spreads outward and results in substantial spatial overlap between both gas and solids.  The sub-Keplerian speed of the gas can
impart an additional (aerodynamic) drag force on the disk solids, increasing their inward drift speed and leading to an enhanced rate of inner gas 
production via sublimation.  Positive feedback results from a rise in the density of gas imparting drag on disk solids, which in turn further amplifies
the sublimation rate.  Successful operation of this runaway mechanism is highly dependent on the coupling strength between gas and dust, but can 
produce accretion rates as high as $10^{11}$\,g\,s$^{-1}$, or a few hundred times those driven by PR drag alone \citep{raf11a,met12}.  

Other methods to produce high-rate accretion require only that a substantial mass of gas is produced in a disk system, comparable to the size of 
a large asteroid.  One possibility is an auxiliary asteroid and its debris impacting the disk from a previous disruption, where roughly the entire mass 
of the smaller body should become gaseous due to high-velocity collisions between debris streams with distinct orbital inclinations \citep{jur08}.  A 
second scenario already discussed is gas produced by collisions during the earliest stages of disk formation, on the orbits immediately following the 
destruction of a single asteroid \citep{far12b,bea13}.  In either case, if sufficient mass in a fully ionized disk of gas results (so that the MRI operates),
then standard $\alpha$ disk accretion \citep{kin07} will result, where rates can easily exceed $10^{11}$\,g\,s$^{-1}$ by a few orders of magnitude.

In addition to the difference in predicted accretion rates between runaway accretion, and gas-dominated accretion model, there is another key
distinction.  Runaway accretion proceeds initially via PR, gradually increasing toward an exponential rise, and ending in a short-lived spike that 
rapidly depletes all the remaining disk, including the bulk of its initial mass \citep{met12}.  Therefore the accretion burst comes at the end of the disk
lifetime, and consumes it utterly.  In contrast, a gas disk produced immediately following a tidal disruption event comes at the beginning of the disk
lifetime (or middle if a secondary impact), and for at least some fraction of cases, leaves behind a dust disk to evolve via PR drag.  Figure \ref{fig10} 
shows there are several helium-rich stars with super-PR accretion rates ($\dot M_{\rm z} > 10^9$\,g\,s$^{-1}$) that still retain visible dust disks, 
and thus the data favor recent but prior events that allow for persistent debris.  It may also be noteworthy that all four DAZ stars with detected 
(overlapping) dust and gas have ongoing accretion rate consistent with PR drag.  However, if multiple events occur within a single diffusion 
timescale for the DBZ-type stars \citep{wya14}, then runaway accretion may operate between planetesimal disruptions.

\subsection{Disk Diversity and Transients}

The above sections present a simplified picture of a single disrupted body of substantial mass to account for the majority of polluted white dwarf 
observations.  But in actuality, it is likely that white dwarf disks are rather varied, and similar in some respects to the planetary rings of the Solar 
System; such diversity and dynamical activity are in fact empirically supported.  Dusty rings with large inner holes, gas emission and variability, 
infrared non-detections and variability, etc.\ all argue for transient phases and active shaping processes.  It is often the case that anomalies and 
outliers provide major leaps of scientific understanding.

Consider the planetary rings of the Solar System, which lie predominantly within the Roche limit of their host.  They display a stunning variety in
size, optical depth, and textures, with moons playing fundamental roles as parent bodies, partitions, sculptors, collectors and dispersers of ring 
material \citep{esp02}.  Rings are active environments, where objects are ground down by (micro-)meteorite impacts, producing fine dust and 
plasma \citep{mit06}, and can harbor transient moonlets -- caught in a balance between fragmentation and accretion -- up to about 1\,km in 
size \citep{esp08}.  Although such detailed phenomena are difficult or impossible to witness in the analogous rings orbiting white dwarfs, the 
processes and structures of planetary rings can provide a basic theoretical framework.  Beyond any compositional differences between their 
ring constituents, planetary and white dwarf rings are mainly dissimilar in the depth of their gravitational wells (i.e.\ higher velocities and shorter 
timescales), and stellar irradiation.

Based on infrared data, the radial extent of the dust rings orbiting white dwarfs is highly varied, and not an observational effect induced by low 
inclination \citep{roc15}.  As discussed above, radiative forces sculpt the outer edges of circumstellar dust and drive those radii inward, but there 
are at least three infrared excesses that manifest only beyond 3.6\,$\mu$m (G166-58, PG\,1225--079, HS\,2132+0941 \citealt{far08a,far10b,
ber14}), not to mention that only about half of all infrared disk detections become apparent at $K$ band.   Because the shortest wavelength at
which thermal emission arises is strongly determined by dust temperature, these cases correspond to different physical disk properties, and not 
their radiative environments (which scale with $T_{\rm eff}$).  Thus, at least three disks are relatively depleted in dust at inner radii where 
sublimation is unimportant, yet retain higher (presumably optically thick) dust densities further out.

Perhaps stronger evidence of disk diversity is the utter lack of infrared detection for many stars where accretion is ongoing \citep{koe14,roc15}.
Possible configurations that would escape detection include the following, some of which can be applied to detected disks with dust depleted 
inner regions: 1) completely gaseous disks \citep{jur08}; 2) extremely narrow but opaque rings \citep{far10b}; 3) optically thin dust \citep{boc11}.
In the latter two scenarios, the disk (segment) escapes infrared detection due to insufficient surface area of the emitting grains, especially in the 
case of optically thin material \citep{deb11b}.  

While there is little doubt that disk solids can easily be vaporized by auxiliary impacts due to different orbital parameters, it is unclear if multiple 
events are compatible in both mass and frequency to remove the bulk of dusty regions that are otherwise detectable.  But more concretely the
ongoing accretion rates determined for warm DAZ stars without infrared excess can be compared to predictions for the three above scenarios.
Gas disks should dissipate rapidly and result in high accretion rates \citep{met12}, possibly orders of magnitude larger than observed to date 
\citep{far12b}, while optically thin or dense but narrow rings will be accreted at modest rates \citep{boc11} compatible with the non-detections 
in Figure \ref{fig10}.  Furthermore, the warm DA stars studied by COS are consistent with a picture where dust often accretes at modest rates, 
and completely gaseous disks dissipate rapidly \citep{koe14}.  Cumulative infrared data suggest disks detections become more frequent as their 
fractional luminosity decreases \citep{roc15}, and thus the body of evidence suggests circumstellar dust is likely present but often undetectable.

In contrast, there are at least two cases where spectacularly strong infrared emission cannot be explained by flat dust disks alone (GD\,56, 
GD\,362; \citealt{jur07a,jur07b}).  In these instances, warped or flared disk segments have been invoked to increase the emitting surface area, 
under the assumption that they are radiatively induced \citep{jur09a}.  These models account for the observations well, but raises the question
of why only these two disks suffer from radiative warping, but no others that orbit stars of similar luminosity.  In analogy with planetary rings, it
seems more straightforward to cause vertical disk deviations using gravity, and thus requiring large fragments equivalent to ring moons.  As an
example, Daphnis has a diameter of 8\,km and causes 1\,km vertical disturbances in the A ring of Saturn; this is 100 times larger than the height 
of the unperturbed rings \citep{wei09}.  Somewhat speculatively, these two white dwarfs may represent particularly active or young ring systems 
with persistent, large fragments orbiting relatively near to their dusty segments.

The role of gas is clearly important, yet there is significant uncertainty in its production and removal from polluted white dwarf systems.  Since all
disks that exhibit gaseous emission components are relatively bright in the infrared, it seems evident that gas production is linked to the presence
of significant solid matter.  On the other hand, although sublimation should be occurring in all dusty systems and is predicted to spread to the dusty
regions \citep{raf11b}, the rare nature of systems with emission features suggests a particular quality such as total disk mass or youth.  An impact
on a pre-existing disk could enhance the mass of vaporized debris, but calculations suggest that any gas co-orbital with an opaque disk should 
rapidly condense, becoming essentially removed within several hundred orbits \citep{met12}.  If correct, this implies that sufficient gas must be 
produced on similar timescales in order to be observed.

All gas emitting disks may display variability in either the strength or shape of their lines, but detailed and long-term monitoring data are only 
just being published \citep{man15}.  SDSS\,1617 had emission features that gradually faded over several years after their onset \citep{wil14}, 
and is consistent with either condensation or $\alpha$ disk accretion.  However, the latter should be accompanied by an increased accretion rate 
(via atmospheric metal abundance) which was not observed.  Given the significant dusty components orbiting all stars with Ca\,{\sc ii} emission, 
gas accretion by the MRI may be suppressed due to small dust grains lowering the ionization fraction \citep{met12}.

SDSS\,1228 has now been monitored for a dozen years and shows intriguing variations in the (generally asymmetric) shape of its emission lines, 
but no significant change in their overall strength.  This prototype and brightest example of a debris disk with gaseous emission appears to have 
a quasi-stable, eccentric gas component that is precessing with a period of a few decades \citep{man15}.  Due to the eccentricity of the gas disk,
it has been hypothesized the entire disk is in the process of circularization and thus continually producing gas via collisions.  For orbital periods
of a few hours at most, this would mean circularization has not occurred within 20\,000--30\,000 orbits where there is substantial gas mass and 
thus viscosity.  While speculative, a physical mechanism may be required to sustain the eccentric structure, and a large fragment analogous to 
a planetary ring moon could have the necessary gravitational influence.  Such a large body would also be fully exposed above the disk plane,
becoming significantly hotter than surrounding (shadowed) debris, more prone to sublimation, and thus a possible source for the gas itself.
 
Lastly and somewhat surprisingly, at least one dusty white dwarf has been shown to vary in the infrared, and interestingly it is also a disk where
Ca\,{\sc ii} emission is detected.  SDSS\,0959 exhibited a drop in its infrared flux by approximately 35\% over a period within 300\,dy, where the
flat disk model projects an increase in the size of the dust depleted, inner regions \citep{xu14}.  Hypotheses include an impact on this inner region
and a disk instability, and while both are predicted to result in rapid accretion, there has been no change observed yet in either the metal absorption 
or emission features (B. T.\ G\"ansicke 2015, private communication); however the precise epoch of the infrared decrease may have been prior to 
the first spectroscopic dataset \citep{far12a}.  Accretion of any vaporized disk mass may be avoided if re-condensation occurs after spreading, or if 
the gas remains stable due to low ionization.  While the wide variety of disk features and phenomena are not yet fully understood, it is increasingly 
likely that the disks with gas emission are particularly active, and processing recent events that induce variability.

\section{Summary and Outlook}

Empirical and theoretical work support a picture where dynamically active planetary systems persist at a significant fraction of all white dwarfs,
clearly manifested by atmospheric metals accreted from circumstellar debris disks.  Key points in the development of this field include:

\begin{itemize}

\item{{\bf Historical firsts} were achieved via observations of polluted and dusty white dwarfs.  The 1917 spectrum of vMa\,2 is the first evidence
collected by astronomers that (with hindsight) indicates the existence of exoplanetary systems.  It is also likely that $K$-band photometry of G29-38 
represents the first detection of exoplanetary material, circa 1980.}

\item{{\bf Metal pollution} describes the presence of atmospheric metals in $T_{\rm eff}\lesssim25\,000$\,K white dwarfs.  The outer layers of these
dense stars should be free of heavy elements, and thus require an external -- and often continuous -- source of infalling metals.  Evidence is now
compelling that circumstellar disks of planetesimal debris are responsible.}

\item{{\bf Infrared observations} have unambiguously connected white dwarf metal pollution and debris disks.  Observations and models indicate
the disks are flat and opaque, extending from a radius where sublimation is efficient to near the Roche limit.  Silicate emission features are detected 
in all observed systems, indicating the presence of micron-size grains.}

\item{{\bf Tidal disruption} of large asteroids is the standard model for these systems.  Debris is not detected exterior to the Roche limit, suggesting
that disks form rapidly from eccentric parent bodies prone to catastrophic fragmentation by gravitational tides.  Theoretical modeling of the initial disk 
formation can constrain important parameters such as total masses and primordial orbital radii.}

\item{{\bf Planetary systems} likely persist at these white dwarfs.  The perturbation of large asteroids is consistent with at least one major planet, but 
retaining sufficient planetesimal belt masses in the (former) terrestrial zone is an outstanding problem.  Kuiper belt analogs can preserve ample mass 
distributions post-main sequence, but to date their volatile-rich chemistry is not observed.}

\item{{\bf Disk accretion} proceeds primarily through Poynting-Robertson drag.  For opaque disks the rate is relatively high and fixed by geometry,
while for optically thin material, the rates are modest; together these account for all systems where a steady-state accretion is likely (i.e.\ warm DAZ
stars).  The largest inferred accretion rates occur only among time-averaged values for DBZ-type stars and indicate high-rate bursts not yet witnessed.}

\item{{\bf Gaseous emission} features are rare, but an important probe of dynamical processes and a critical real-time view of active planetary
systems.  The production and behavior of detected gas is still somewhat uncertain, but monitoring this material may reveal clues to disk and parent 
body origins, viscous evolution, auxiliary impacts, etc.}

\item{{\bf Disk diversity} has been increasing in recent years, and it is now clear that the bulk of disks lie below infrared detection thresholds.  The
sizes and shapes inferred from thermal and gaseous emission show compelling evidence for evolution under the influence of white dwarf radiation,
and possibly gravitational forces from additional bodies.}

\end{itemize}

White dwarfs are compelling targets for exoplanetary system research, as their polluted surfaces reveal unique and powerful insight into the chemical
and physical assembly of exoplanets.  This is especially true in the case of asteroid analogs, where metal abundances then provide invaluable empirical 
data for the building blocks of terrestrial exoplanets.  Furthermore, white dwarfs are the descendants of intermediate-mass stars that are typically immune
to conventional exoplanet detection, especially within their inner regions.  The review presented here, and specifically the research within, provides a
broad context to connect origins, formation, and evolution with the remnant yet active systems observed today.

After submission of this review, the first transit events around a white dwarf were reported, based on {\em K2} observations of WD\,1145+107 
\citep{van15}.  The photometric data are fairly compelling, with myriad ground-based follow up observations confirming multiple, broad, complex,
and evolving transits with periods of approximately five hours and thus near to the Roche limit \citep{cro15,gan16}.  Furthermore, WD\,1145+017
appears highly polluted with metals and has an infrared excess in {\em WISE}, and thus at face value it appears to provide striking confirmation 
of the asteroid disruption model.  While it is too early to know what might be learned from the continued monitoring of this system, it is likely to
provide unprecedented insight into disk formation and evolution.  Perhaps most interestingly for the bigger picture, it remains to be seen if the
detection of these transits is most dependent on geometry or timing (or both).

The near future holds some outstanding prospects for progressing the understanding of disks around white dwarfs.  Currently there are basically
no empirical constraints on the parent bodies themselves, apart from their bulk chemical composition.  The best understanding of likely source
regions and sizes comes primarily from theoretical considerations \citep{bon11,wya14}, and if additional transiting systems are found they may 
help to independently constrain the sizes and orbits of persistent planetesimals.  Direct or indirect detection of the parent planetesimal belts is 
possible in the sub-millimeter with ALMA or with future far-infrared space missions, but may require a substantial investment to identify favorable
systems \citep{far14}.  In-situ dust mineralogy is possible with {\em JWST} and should be highly complementary to the bulk abundances revealed
from atmospheric pollution.  Lastly, the detection of giant planets in these systems may be possible with {\em JWST}, but if any persist within 
several AU, they will not escape astrometric detection by {\em GAIA}. 

\section*{Acknowledgments}

Prof.\ M.\ Jura (1947--2016) had an extraordinary influence on the content of this review, from both observational and theoretical perspectives.  His many 
pioneering contributions to this emerging field cannot be underestimated, and his untimely death is a major loss to the scientific community.  The author 
is deeply grateful for many discussions and collaborations with Prof.\ Jura over the years, and acknowledges an enormous debt to him as a scientist.

The author thanks several colleagues for useful discussions and feedback on a draft manuscript: J. H. Debes, B. T. G\"ansicke, R. R. Rafikov, D. Veras, 
and M. C. Wyatt.  An anonymous referee made suggestions that significantly improved this review, and for which the author is grateful.  Special thanks go 
to H. A. McAlister and J. S. Mulchaey for their assistance in tracking down the 1917 photographic plate with the spectrum of vMa\,2.   J. Farihi is supported 
by the UK Science and Technology Facilities Council in the form of an Ernest Rutherford Fellowship.

\end{document}